\documentclass[extra,mreferee]{gji}
\usepackage{timet,color}
\usepackage{graphicx}
\usepackage{lineno}
\usepackage{amsmath}
\usepackage{textcomp}
\usepackage{multicol}
\usepackage{multirow}
\usepackage{hyperref}

\title[MFSM for calculating t*]
{A novel fast sweeping method for computing the attenuation operator $t^*$ in absorbing media}

\author[Wang et al., 2024]
  {Dongdong Wang$^1$, Jing Chen$^1$, Shijie Hao$^1$, and Ping Tong$^{1,2,3}$ \\
  $^1$ Division of Mathematical Sciences, School of Physical and Mathematical Sciences, Nanyang Technological \\ University, Singapore, Singapore. E-mail: \href{tongping@ntu.edu.sg}{tongping@ntu.edu.sg} \\
  $^2$ Earth Observatory of Singapore, Nanyang Technological University, Singapore Singapore\\
  $^3$ Asian School of the Environment, Nanyang Technological University, Singapore, Singapore
  }

\begin{document}

\maketitle

\begin{summary}
$t^*$ represents the total path attenuation and characterizes the amplitude decay of a propagating seismic wave. Calculating the attenuation operator $t^*$ is typically required in seismic attenuation tomography. Traditional methods for calculating $t^*$ require determining the ray path explicitly. However, ray tracing can be computationally intensive when processing large datasets, and conventional ray tracing techniques may fail even in mildly heterogeneous media. In this study, we propose a modified fast sweeping method (MFSM) to solve the governing equation for $t^*$ without explicitly calculating the ray path. The approach consists of two main steps. First, the traveltime field is calculated by numerically solving the eikonal equation using the fast sweeping method. Second, $t^*$ is computed by solving its governing equation with the MFSM, based on the discretization of the gradient of $t^*$ using an upwinding scheme derived from the traveltime gradient. The MFSM is rigorously validated through comparisons with analytical solutions and by examining $t^*$ errors under grid refinement in both simple and complex models. Key performance metrics, including convergence, number of iterations, and computation time, are evaluated. Two versions of the MFSM are developed for both Cartesian and spherical coordinate systems. We demonstrate the practical applicability of the developed MFSM in calculating $t^*$ in North Island, and discuss the method's efficiency in estimating earthquake response spectra.





\end{summary}


\section{Introduction}\label{intro}




Seismic attenuation, apart from geometrical spreading, is the primary process that reduces the amplitude and modifies the phase of a propagating seismic wave. Attenuation is quantified by the quality factor, $Q$, defined as the ratio of energy lost during a wave cycle to the total energy of the cycle. The reciprocal of $Q$, denoted as $1/Q$, depends on the rock properties and accounts for energy loss through elastic and anelastic mechanisms, referred to as scattering and intrinsic attenuation, respectively \citep{sato2012}. Scattering attenuation arises from the interaction of seismic waves with small-scale heterogeneities in the elastic properties of the medium, such as those caused by intense rock fracturing. On the other hand, intrinsic attenuation occurs when the kinetic energy of seismic waves is converted into thermal energy, either through internal friction along cracks or via viscoelastic deformation of the medium. In practical applications, with separation of seismic scattering from absorption, the inversion yields intrinsic $Q$; without separation, the inversion yields total $Q$.



A widely used method for determining $Q$ is based on the estimation of the attenuation operator $t^*$, which accounts for the damping of the wave amplitude $A$ through the exponential decay $e^{-{{\omega}{t^*}/2}}$ \citep[e.g.,][]{cormier1982,bindi2006}, where $\omega$ is the angular frequency. Physically, $t^*$ represents the cumulative attenuation along a ray path connecting the hypocentre to the station \citep{kanamori1967}. It can be mathematically defined as $t^* =\int_{L}{1}/({V(\boldsymbol{x})Q(\boldsymbol{x}))}dl$, where $V(\boldsymbol{x})$ and $Q(\boldsymbol{x})$ are the seismic velocity and quality factor, respectively, $L$ is the ray path. Synthetic $t^*$ values are typically calculated by summation along the ray path \citep[e.g.,][]{lees1994,eberhart2008,desiena2009}. Consequently, the accuracy of $t^*$ is heavily dependent on the accuracy of the ray path. Over the past few decades, a variety of seismic ray tracing techniques have been developed to determine ray paths. These include shooting and bending ray tracing methods, as well as numerical solutions to the eikonal equation on a grid. Specifically, shooting methods treat the ray equation as an initial value problem, iteratively adjusting the ray's take-off angle until the source-receiver path is found \citep[e.g.,][]{cerveny1987,sambridge1990,rawlinson2001}. Bending methods \citep[e.g.,][]{julian1977,um1987} iteratively modify the geometry of an initially assumed path between the source and receiver until it conforms to the Fermat's principle. The pseudo-bending technique \citep{um1987} and the thurber-modified ray-bending approach \citep{block1991}, have been widely used to trace ray paths in attenuation tomography \citep[e.g.,][]{lees1994,eberhart2008,wei2018,desiena2009,desiena2010,desiena2014,prudencio2015anta,sketsiou2021}.


However, both shooting and bending methods may fail to converge on the true ray paths in the presence of velocity variations \cite[e.g.][]{rawlinson2004}. This issue becomes increasingly pronounced as the complexity of the medium grows. Recently, grid-based schemes, such as the fast marching method (FMM) \citep[e.g.,][]{sethian1996,sethian1999,alkhalifah2001} and the fast sweeping method (FSM) \citep[e.g.,][]{zhao2005fast,qian2007fast,luo2012}, have gained significant popularity. These methods numerically solve the eikonal equation on a gridded velocity field to compute the traveltime from the source to every grid point. The ray path is then traced from the receiver to the source along the negative gradient of the traveltime field. These approaches are fast, accurate, and robust for calculating traveltime fields, even in complex heterogeneous media \citep{rawlinson2010}. Theoretically, grid-based methods can provide relatively accurate ray paths, enabling the summation along these paths to yield a more precise estimation of $t^*$.

The aforementioned studies focus on methods for calculating $t^*$ that rely on ray tracing. However, ray tracing can be computationally expensive, particularly when dealing with a large number of sources and/or receivers in a 3-D medium \citep{rawlinson2004}. In addition, the development of the adjoint-state attenuation tomography method requires the calculation of $t^*$ without the use of ray tracing \citep[e.g.,][]{huang2020,he2020}. To address this issue and take advantages of grid-based methods, a grid-based approach for calculating $t^*$ needs to be developed. By converting the integral form of $t^*$ to its differential form and considering the relationship between the gradient and the directional derivative, \cite{huang2020} derive the governing equation for $t^*$. Later, \cite{he2022} reformulates the governing equation for $t^*$ using the Leibniz formula and develops a parallel FSM to solve it. However, their study would benefit from further verification and evaluation, as well as a more in-depth analysis of accuracy and convergence, and is primarily focused on solving the equation in Cartesian coordinates.

In this paper, we develop a modified fast sweeping method (MFSM) as an alternative approach for solving the $t^*$ governing equation. Considering that $t^*$ accumulates along the ray path, which is not directly determined by traveltime but rather by the traveltime gradient, the calculation of $t^*$ thus depends on the traveltime gradient. The traveltime field is first calculated using the FSM, after which $t^*$ is determined by the MFSM through an upwinding scheme derived from the traveltime gradient. Our proposed MFSM provides an effective and accurate method for calculating $t^*$ without the need for ray tracing and serves as a promising forward modeling tool for adjoint-state attenuation tomography.

The paper is organized as follows. From the definition of $t^*$, we provide a complete derivation of the governing equation for $t^*$ in differential form in Section \ref{sec_2_1}. Then, the FSM for solving the eikonal equation is reviewed in Section \ref{sec_2_2}, and the MFSM for solving the $t^*$ governing equation is introduced in Section \ref{sec_2_3}. In Section \ref{sec_3}, considering several simple and complex models, we verify the MFSM by comparing its solutions with the analytical solutions and analyze the relative and absolute $t^*$ errors with grid refinement. The convergence analysis, number of iterations, and computational time are presented in Section \ref{sec_4}. In Section \ref{sec_5}, we consider realistic velocity and attenuation models for the central North New Zealand region and apply the MFSM to calculate and analyze the $t^*$ field. Finally, we discuss and summarize the results in Section \ref{sec_6}.


\section{Governing equations and numerical algorithms}\label{sec_2}

\subsection{Governing equation for the attenuation operator $t^{*}$}\label{sec_2_1}
We denote the propagation velocity of the wave under consideration (either $P$ or $S$) by $V(\boldsymbol{x})$. The traveltime $t$ from the source to the receiver can be calculated by integrating the inverse of the velocity along the ray path connecting them \citep{cerveny2001}
\begin{equation}\label{t_path}
t =\int_{L} \frac{1}{V(\boldsymbol{x})}dl,
\end{equation}
where $dl$ is the arclength along the ray path $L$. In viscoelastic media, the inverse of the quality factor, $1/Q$, is commonly used to quantify attenuation. A high $Q$ indicates low attenuation, while a low $Q$ indicates high attenuation. Previous studies suggest that attenuation primarily affects the waveform through the complex and frequency-dependent traveltime, rather than considerably altering ray paths, provided that $1/Q \ll 1$ \citep{keers2001}. The attenuation operator $t^{*}$ can be expressed as \citep[e.g.,][]{stachnik2004, wei2018}
\begin{equation}\label{tstar_path}
t^* =\int_{L}\frac{1}{V(\boldsymbol{x})Q(\boldsymbol{x})}dl,
\end{equation}
where $Q(\boldsymbol{x})$ represents the P-wave or S-wave quality factor. The integral eqs \eqref{t_path} and \eqref{tstar_path} can be rewritten in differential forms as,
\begin{equation}\label{t_dif}
\frac{dt}{dl} = \frac{1}{V(\boldsymbol{x})},
\end{equation}
\begin{equation}\label{tstar_dif}
\frac{d{t^*}}{dl} = \frac{1}{V(\boldsymbol{x})Q(\boldsymbol{x})},
\end{equation}
where both ${dt}/{dl}$ and ${d{t^*}}/{dl}$ represent the directional derivatives along the ray paths. Mathematically, these two differentials can be expressed as:
\begin{equation}\label{t_grad}
\frac{dt}{dl} = \nabla t (\boldsymbol{x}) \cdot \frac{\boldsymbol{l}}{\lvert {\boldsymbol{l}} \rvert},
\end{equation}
\begin{equation}\label{tstar_grad}
\frac{d{t^*}}{dl} = \nabla t^* (\boldsymbol{x}) \cdot \frac{\boldsymbol{l}}{\lvert {\boldsymbol{l}} \rvert},
\end{equation}
where ${\boldsymbol{l}}/{\lvert {\boldsymbol{l}} \rvert}$ denotes the unit tangent vector along the ray path direction, and ${\nabla t}$ and ${\nabla t^*}$ represent the gradients of $t$ and $t^*$, respectively. Notably, the same unit vector ${\boldsymbol{l}}/{\lvert {\boldsymbol{l}} \rvert}$ appearing in eqs \eqref{t_grad} and \eqref{tstar_grad} indicates the identical evolution direction for both $t$ and $t^*$. In an isotropic medium, the unit tangent vector ${\boldsymbol{l}}/{\lvert {\boldsymbol{l}} \rvert}$ is equivalent to the unit normal of the wavefront:
\begin{equation}\label{direction}
\frac{\boldsymbol{l}}{\lvert {\boldsymbol{l}} \rvert}=\frac{\nabla t (\boldsymbol{x})}{\lvert \nabla t (\boldsymbol{x}) \rvert}.
\end{equation}
For simplicity, in the following derivation we use $s(\boldsymbol{x})$ (slowness) to represent $1/V(\boldsymbol{x})$ and $q(\boldsymbol{x})$ to represent $1/Q(\boldsymbol{x})$. Based on eqs \eqref{t_dif}, \eqref{t_grad}, \eqref{direction}, the eikonal equation governing wavefront propagation from the point source $\boldsymbol{x}_s$ to any position $\boldsymbol{x}$, with a zero boundary condition at $\boldsymbol{x}_s$, can be expressed as
\begin{equation}\label{t_eik}
\nabla t (\boldsymbol{x})  \cdot \nabla t (\boldsymbol{x}) = s^2 (\boldsymbol{x}), \quad t(\boldsymbol{x}_{s}) = 0.
\end{equation}
It is important to note that under the high-frequency approximation, eq. $\eqref{t_eik}$ can also be derived from the wave equation \citep{shearer2019}. Similarly, combining eqs \eqref{tstar_dif}, \eqref{tstar_grad}, \eqref{direction} and \eqref{t_eik}, we can obtain
\begin{equation}\label{tstar_eik}
\nabla t (\boldsymbol{x})  \cdot \nabla t^* (\boldsymbol{x}) = q(\boldsymbol{x})s^2 (\boldsymbol{x}), \quad t^*(\boldsymbol{x}_{s}) = 0.
\end{equation}
Thus, we have established eq. \eqref{tstar_eik} as the governing equation for $t^*$ in attenuation media. The derivation of eq. \eqref{tstar_eik} follows a similar workflow to that presented in \cite{huang2020}, but we provide a more detailed procedure here. One difference is that, unlike eq. \ref{t_path}, in \cite{huang2020}, $t^*$ is defined as $t^* =\int_{L}{\pi}/({V(\boldsymbol{x})Q(\boldsymbol{x}))}dl$. Thus, the constant $\pi$ is absent from the right-hand side of eq. \ref{tstar_eik} compared with that in \cite{huang2020}. Both types of $t^*$ governing equation are practical, as we can choose to include the constant $\pi$ in the governing equation for $t^*$, or incorporate it when measuring the observed $t^*$. In the following, we will introduce how to solve eq. \eqref{tstar_eik} using the fast sweeping method (FSM). 


\begin{figure}
\center{\includegraphics[width=15cm]{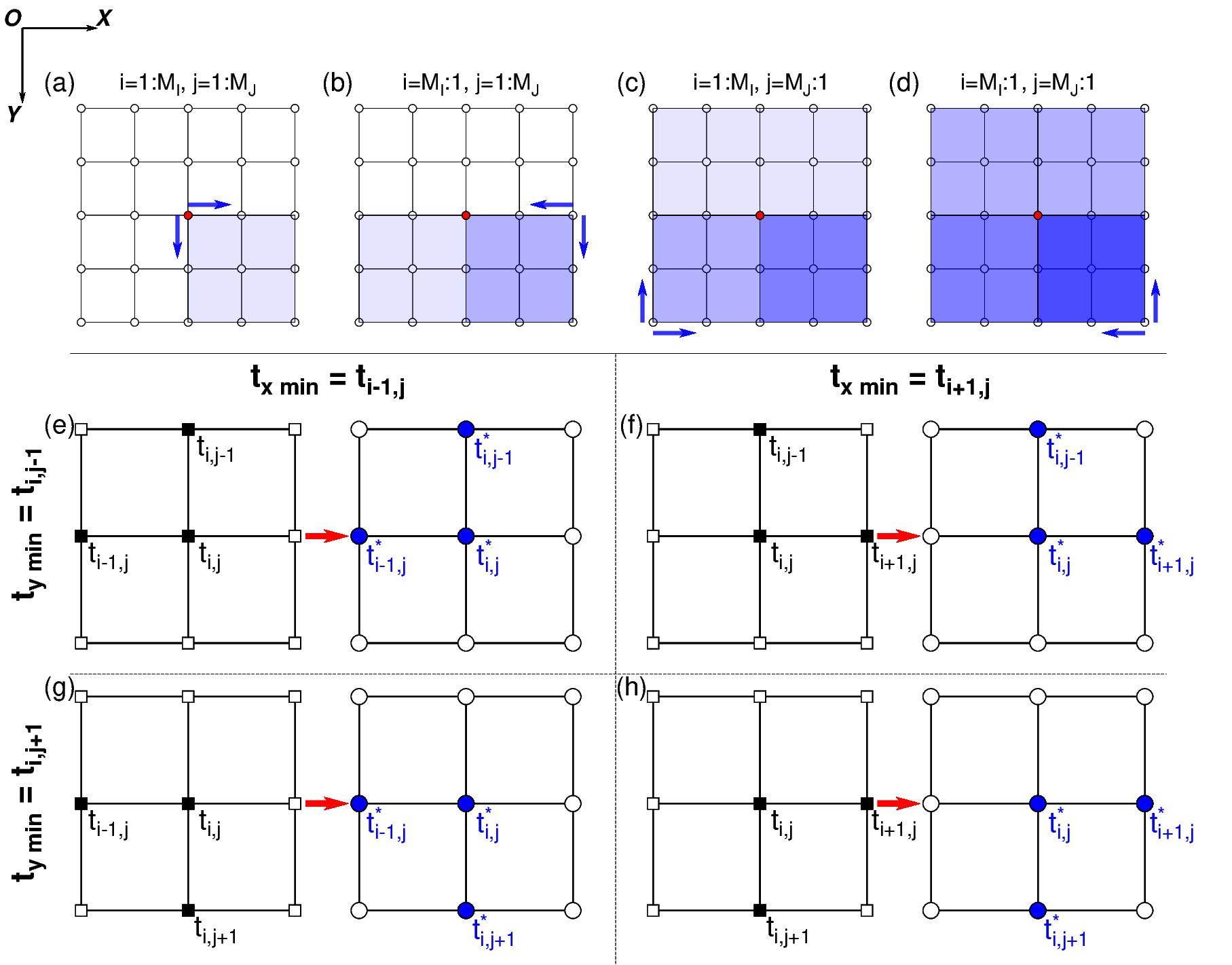}}
\caption
{Illustration of the sweeping order in the fast sweeping algorithm and the discretization of partial derivatives of $t^*$ using the upwind difference scheme for $t$ in Cartesian coordinates. In (a), (b), (c), and (d), the blue arrows represent the sweeping directions, starting respectively from the upper-left, upper-right, lower-left, and lower-right corners of the computational domain. The red circle marks the source location. Nodes within the white area represent grid points where $t$ (or $t^*$) values are yet to be updated, while nodes in the blue area indicate grid points where $t$ (or $t^*$) values have already been updated. Darker shades of blue signify more recent updates. (e), (f), (g), and (h) display four conditions for selecting grids to discretize the partial derivatives of $t^*$ using the upwind difference scheme for $t$. Black squares denote traveltime $t$, while blue circles represent $t^*$.}
\label{fig:fsm}
\end{figure}

\subsection{Overview of the FSM for solving the eikonal eq. \eqref{t_eik}}\label{sec_2_2}

To compute $t^{*}(\boldsymbol{x})$ using eq. \eqref{tstar_eik}, it is first necessary to determine the traveltime field $t(\boldsymbol{x})$ by solving eq. \eqref{t_eik}. In this study, we use the fast sweeping method (FSM), an iterative algorithm, to solve eq. \eqref{t_eik} on a rectangular grid to obtain the traveltime field $t(\boldsymbol{x})$ \citep{zhao2005fast}. Below, we provide a brief overview of the FSM for solving eq. \eqref{t_eik}. The core idea of the FSM is the utilization of nonlinear upwind differences combined with Gauss-Seidel iterations, executed in alternating sweeping orders. This method is straightforward to implement and highly efficient for parallel computation.


We present the algorithm for solving eq. \eqref{t_eik} in 3D Cartesian coordinates \citep{zhao2005fast}. To address applications on a global scale, we also extend the algorithm to spherical coordinates, as detailed in Appendix A. In 3D Cartesian coordinates ($\boldsymbol{x}=(x,y,z)$), the gradient operator is expressed as $\nabla t(\boldsymbol{x})=(\partial_{x}t, \partial_{y}t, \partial_{z}t)$. First, the 3D domain $\Omega \subset \mathcal{R}^3$ is discretized into a uniform mesh of grid points $\boldsymbol{x}_{i,j,k}$ with grid spacings $\Delta x$, $\Delta y$, and $\Delta z$. The total number of grid points in the $x$, $y$, and $z$ directions are $M_{I}$, $M_{J}$, and $M_{K}$. Employing the Godunov upwind difference scheme to discretize eq. \eqref{t_eik} at interior grid points $(2 \leq i \leq M_{I}-1,2 \leq j \leq M_{J}-1,2 \leq k \leq M_{K}-1)$, we obtain
\begin{equation}\label{t_disc}
[\frac{(t_{i,j,k}-t_{i,j,k}^{x\, \mathrm{min}})^+}{\Delta x}]^2+
[\frac{(t_{i,j,k}-t_{i,j,k}^{y\, \mathrm{min}})^+}{\Delta y}]^2 + 
[\frac{(t_{i,j,k}-t_{i,j,k}^{z\, \mathrm{min}})^+}{\Delta z}]^2 
= s_{i,j,k}^2,
\end{equation}
where
\begin{equation}\label{t_min}
t_{i,j,k}^{x\, \mathrm{min}} = \mathrm{min}(t_{i-1,j,k},t_{i+1,j,k}),\,\,t_{i,j,k}^{y\, \mathrm{min}} = \mathrm{min}(t_{i,j-1,k},t_{i,j+1,k}),\,\,t_{i,j,k}^{z\, \mathrm{min}} = \mathrm{min}(t_{i,j,k-1},t_{i,j,k+1}),
\end{equation}
and
\begin{equation}\label{x_plus}\begin{split}
(x)^+ = \left\{
      \begin{aligned}
      x, \: x \enspace \textgreater 0,\\
      0, \: x \leq 0.
      \end{aligned}
\right.
\end{split}\end{equation}
At the boundaries of the region (i.e., $i=1\lor M_{I}$; $j=1\lor M_{J}$; $k=1\lor M_{K}$), one-sided difference schemes are applied. For instance, at the left boundary $\boldsymbol{x}_{1,j,k}$, the use of a one-sided difference yields
\begin{equation}\label{t_one_side}
[\frac{(t_{1,j,k}-t_{2,j,k})^+}{\Delta x}]^2+
[\frac{(t_{1,j,k}-t_{1,j,k}^{y\, \mathrm{min}})^+}{\Delta y}]^2 + 
[\frac{(t_{1,j,k}-t_{1,j,k}^{z\, \mathrm{min}})^+}{\Delta z}]^2 
= s_{1,j,k}^2,
\end{equation}

Subsequently, eqs \eqref{t_disc} and \eqref{t_one_side} can be solved using the Fast Sweeping Algorithm \citep[e.g.,][]{zhao2005fast,leung2006}:

(i) \textbf{Initialization}: If the point source is located exactly at a grid point, the traveltime $t$ is set as $t(\boldsymbol{x}_{s}) = 0$. If the point source within the 3D mesh does not coincide with any grid points, the traveltimes from the source to adjacent grid points are initiated by multiplying the distance from the source to each adjacent grid point by a constant slowness. These values remain fixed during subsequent iterations. Sufficiently large positive values are assigned to all other grid points, which will be updated in later iterations.

(ii) \textbf{Iteration}: Gauss-Seidel iterations with alternating sweeping orders are used to update the solution. The candidate traveltime $\overline{t}$ at each grid point is computed by solving eqs \eqref{t_disc} and \eqref{t_one_side} based on the current values of its neighbouring grid points $t_{i\pm 1,j,k}$, $t_{i,j\pm 1,k}$, $t_{i,j,k \pm 1}$. Then, $t_{i,j,k}$ is updated by selecting the smaller value between its current value $t^{old}_{i,j,k}$ and the candidate $\overline{t}$, i.e., $t^{new}_{i,j,k} = \mathrm{min}(t^{old},\overline{t})$. Figs \ref{fig:fsm}(a)–(d) illustrate the sweeping order in 2D Cartesian coordinates. In 3D Cartesian coordinates, the entire region is swept in the following orders:

(1) $i=1:M_{I}, j=1:M_{J}, k=1:M_{K}$;  (2) $i=M_{I}:1, j=1:M_{J}, k=1:M_{K}$;

(3) $i=1:M_{I}, j=M_{J}:1, k=1:M_{K}$;  (4) $i=M_{I}:1, j=M_{J}:1, k=1:M_{K}$;

(5) $i=1:M_{I}, j=1:M_{J}, k=M_{K}:1$;  (6) $i=M_{I}:1, j=1:M_{J}, k=M_{K}:1$;

(7) $i=1:M_{I}, j=M_{J}:1, k=M_{K}:1$;  (8) $i=M_{I}:1, j=M_{J}:1, k=M_{K}:1$.

(iii) \textbf{Convergence}: A convergence criterion is set with a parameter $\varepsilon \: \textgreater \: 0$. The algorithm checks whether the traveltime difference between two successive iterations satisfies $|| t^{n+1} - t^{n} ||_{L1} \leq \varepsilon$.

\subsection{Modified FSM for solving the $t^*$ governing eq. \eqref{tstar_eik}}\label{sec_2_3}

In Section \ref{sec_2_2}, we reviewed the FSM method for calculating $t(\boldsymbol{x})$. In this section, we present a method for solving eq. \eqref{tstar_eik} using the well-determined $t(\boldsymbol{x})$, along with the velocity and attenuation models, to compute $t^*(\boldsymbol{x})$. To achieve this, we propose a modified fast sweeping method (MFSM) to calculate $t^*(\boldsymbol{x})$ based on $t(\boldsymbol{x})$ and its gradient.

In 3D Cartesian coordinates, we discretize eq. \eqref{tstar_eik} at the interior grid points $(2 \leq i \leq M_{I}-1,2 \leq j \leq M_{J}-1,2 \leq k \leq M_{K}-1)$ into the following difference form:
\begin{equation}\label{tstar_disc}\begin{split}
&\frac{(t_{i,j,k}-t_{i,j,k}^{x\, \mathrm{min}})^+}{\Delta x}\frac{(t^*_{i,j,k}-t^{*,\, x\, \mathrm{min}}_{i,j,k})}{\Delta x}+
\frac{(t_{i,j,k}-t_{i,j,k}^{y\, \mathrm{min}})^+}{\Delta y}\frac{(t^*_{i,j,k}-t^{*,\,y\, \mathrm{min}}_{i,j,k})}{ \Delta y} +\\& 
\frac{(t_{i,j,k}-t_{i,j,k}^{z\, \mathrm{min}})^+}{\Delta z}\frac{(t^*_{i,j,k}-t^{*,\,z\, \mathrm{min}}_{i,j,k})}{ \Delta z}
= s_{i,j,k}^2 q_{i,j,k},
\end{split}\end{equation}
where $\nabla t(\boldsymbol{x})$ is discretized using the Godunov upwind difference scheme, as is done when solving the eikonal equation. Here, $t_{i,j,k}^{x\, \mathrm{min}}$, $t_{i,j,k}^{y\, \mathrm{min}}$ and $t_{i,j,k}^{z\, \mathrm{min}}$ are determined based on eq. \eqref{t_min}.  Since  $t^*$ accumulates in the same direction as $t$ as the wave propagates outward, and the propagation direction depends on $\nabla t(\boldsymbol{x})$, the approximation of $\nabla t^*$ requires that the selection of $t^{*,\, x\, \mathrm{min}}_{i,j,k}$ aligns with the choice of $t^{x\, \mathrm{min}}_{i,j,k}$, as done when determining  $\nabla t(\boldsymbol{x})$.  Figs \ref{fig:fsm}(e)-(h) illustrate how $t^{*,\,x\, \mathrm{min}}_{i,j}$ and $t^{*,\,y\, \mathrm{min}}_{i,j}$ are determined based on $t_{i,j}^{x\, \mathrm{min}}$ and $t_{i,j}^{y\, \mathrm{min}}$ in 2D Cartesian coordinates. A similar strategy is applied in 3D Cartesian coordinates. Mathematically, $t^{*,\,x\, \mathrm{min}}_{i,j,k}$, $t^{*,\,y\, \mathrm{min}}_{i,j,k}$, and $t^{*,\,z\, \mathrm{min}}_{i,j,k}$ can be determined using the following rulers:
\begin{equation}\label{tstar_xmin}\begin{split}
t^{*,\, x\, \mathrm{min}}_{i,j,k} = \left\{
      \begin{aligned}
      t^*_{i-1,j,k}, \: \enspace \textup{if} \enspace t_{i,j,k}^{x\, \mathrm{min}} = t_{i-1,j,k} ,\\
      t^*_{i+1,j,k}, \: \enspace \textup{if} \enspace {t_{i,j,k}^{x\, \mathrm{min}} = t_{i+1,j,k}} ,
      \end{aligned}
\right.
\end{split}\end{equation}
\begin{equation}\label{tstar_ymin}\begin{split}
t^{*,\, y\, \mathrm{min}}_{i,j,k} = \left\{
      \begin{aligned}
      t^*_{i,j-1,k}, \: \enspace \textup{if} \enspace t_{i,j,k}^{y\, \mathrm{min}}= t_{i,j-\rm{1},k}  ,\\
      t^*_{i,j+1,k}, \: \enspace \textup{if} \enspace t_{i,j,k}^{y\, \mathrm{min}} = t_{i,j+1,k},
      \end{aligned}
\right.
\end{split}\end{equation}
\begin{equation}\label{tstar_zmin}\begin{split}
t^{*,\, z\, \mathrm{min}}_{i,j,k} = \left\{
      \begin{aligned}
      t^*_{i,j,k-1}, \: \enspace \textup{if} \enspace t_{i,j,k}^{z\, \mathrm{min}} = t_{i,j,k-1},\\
      t^*_{i,j,k+1}, \: \enspace \textup{if} \enspace t_{i,j,k}^{z\, \mathrm{min}} = t_{i,j,k+1}.
      \end{aligned}
\right.
\end{split}\end{equation}
It should be noted that, similar to eq. \eqref{t_one_side}, one-sided differences are employed at the boundaries of the computational domain. For instance, at the boundary point $\boldsymbol{x}_{1,j,k}$, we have
\begin{equation}\label{tstar_one_side}\begin{split}
&\frac{(t_{1,j,k}-t_{2,j,k})^+}{\Delta x}\frac{(t^*_{1,j,k}-t^{*}_{2,j,k})}{\Delta x}+
\frac{(t_{1,j,k}-t_{1,j,k}^{y\, \mathrm{min}})^+}{\Delta y}\frac{(t^*_{1,j,k}-t^{*,\,y\, \mathrm{min}}_{1,j,k})}{ \Delta y} +\\& 
\frac{(t_{1,j,k}-t_{1,j,k}^{z\, \mathrm{min}})^+}{\Delta z}\frac{(t^*_{1,j,k}-t^{*,\,z\, \mathrm{min}}_{1,j,k})}{ \Delta z}
= s_{1,j,k}^2 q_{1,j,k}.
\end{split}\end{equation}

Then, eqs \eqref{tstar_disc} and \eqref{tstar_one_side} can be solved using the following Fast Sweeping Algorithm:

(i) \textbf{Initialization}: If the point source is located exactly at a mesh point, $t^*$ is initiated as $t^*(\boldsymbol{x}_{s}) = 0$. If the point source lies with the 3D mesh but not directly on a grid point, $t^*$ at the adjacent grid points is assigned values computed as the product of the distance from the source to the grid point, the square of a constant slowness $s (\boldsymbol{x})$, and a constant $q (\boldsymbol{x})$. These values remain fixed in subsequently iterations. All other mesh points are assigned sufficiently large positive values, which will be updated iteratively.

(ii) \textbf{Iteration}: Similar to the method described in Section \ref{sec_2_2}, Gauss-Seidel iterations with alternating sweeping directions are employed to update the solution. By solving eqs \eqref{tstar_disc} and \eqref{tstar_one_side}, the candidate solutions $\overline{t^*}$ at each grid point is calculated using the neighboring values $t_{i\pm 1,j,k}$, $t_{i,j\pm 1,k}$, $t_{i,j,k \pm 1}$, $t^*_{i\pm 1,j,k}$, $t^*_{i,j\pm 1,k}$ and $t^*_{i,j,k \pm 1}$. The updated value $t^*_{i,j,k}$ is then selected as the minimum between its current value and the candidate solution $\overline{t^*}$, i.e., $t^{*new}_{i,j,k} = \mathrm{min}(t^{*old},\overline{t^*})$. As shown in Figs \ref{fig:fsm}(a)-(d), the sweeping order is repeated in the same sequence as that used in Section \ref{sec_2_2}. It is important to emphasize that the $t (\boldsymbol{x})$ field remains fixed during these sweeps.

(iii) \textbf{Convergence}: The convergence criterion for $t^*$ is set with a parameter $\upsilon$. Convergence is achieved when the difference between $t^*$ at iteration $n+1$ and iteration $n$ satisfies the condition: $|| t^{*n+1} - t^{*n} ||_{L1} \leq \upsilon$.

Thus, we have established the theoretical framework for the Modified Fast Sweeping Method (MFSM) to determine $t^*$ in Cartesian coordinates. The MFSM for determining $t^*$ in spherical coordinates, based on eq. \eqref{tstar_eik}, can be found in Appendix A.

\begin{figure}
\center{\includegraphics[width=13cm]{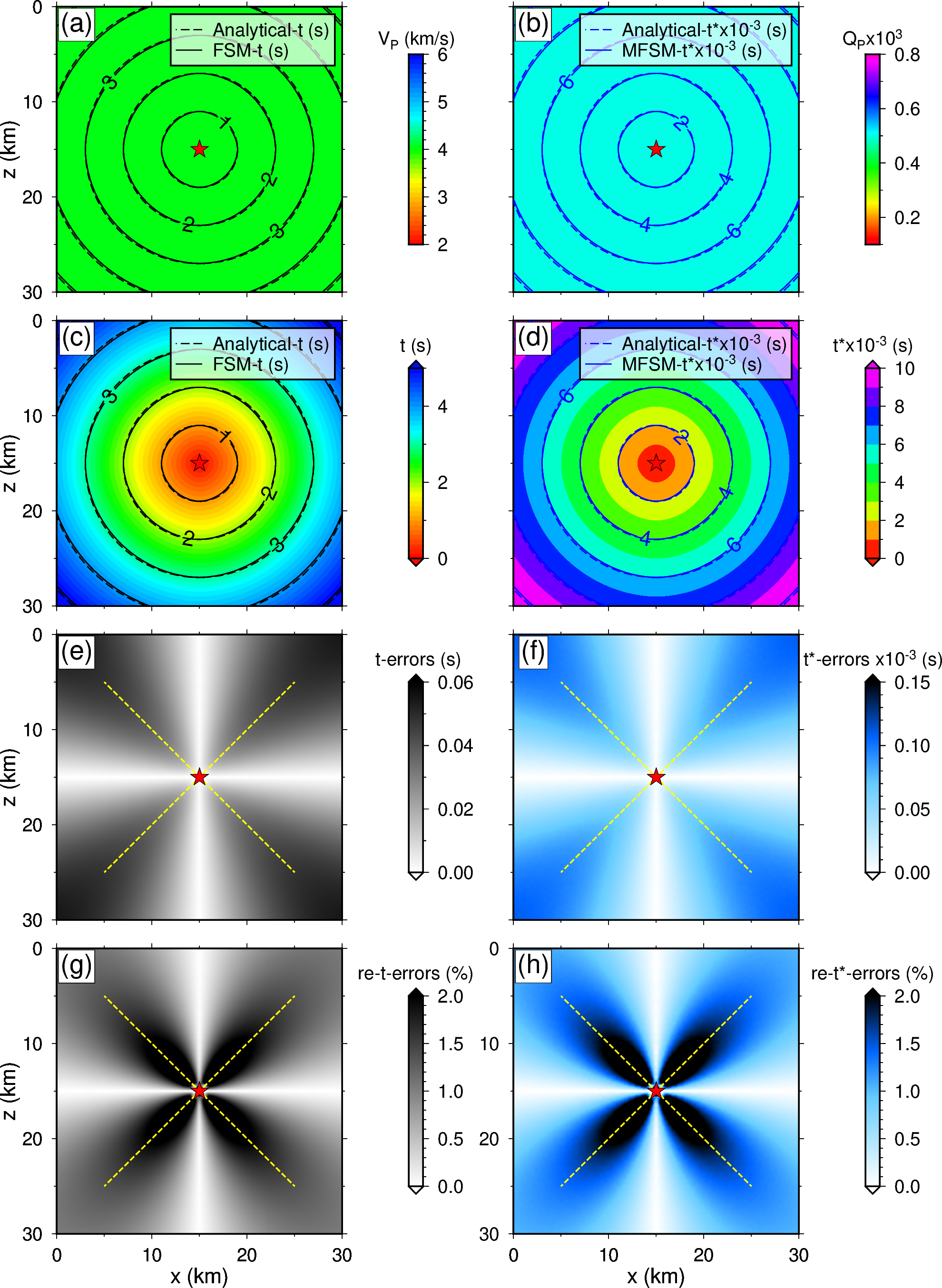}}
\caption
{Model 1 consists of a uniform $V_{P}$ (a) and a uniform $Q_{P}$ (b) in Cartesian coordinates. The $V_{P}$ value in (a) is specified as 4.0 km/s, while the $Q_{P}$ value in (b) is set as 500.0. The source (red star) is positioned at the location (15 km, 15 km). The region is discretized with a 0.20 km $\times$ 0.20 km grid, resulting in 151 $\times$ 151 points. The comparison of isolines calculated using the numerical and analytical solutions for $t(\boldsymbol{x})$ is shown in (a) and (c), while the comparison for $t^*(\boldsymbol{x})$ is shown in (b) and (d), respectively. The $t(\boldsymbol{x})$ and $t^*(\boldsymbol{x})$ fields are shown in (c) and (d), respectively. The absolute and relative (in percentage) errors for $t(\boldsymbol{x})$ are shown in (e) and (g), while the absolute and the relative (in percentage) errors for $t^*(\boldsymbol{x})$ are shown in (f) and (h). The yellow dashed lines represent auxiliary lines along the diagonal directions.}
\label{fig:model1}
\end{figure}

\section{Verification of the MFSM}\label{sec_3}

In Section \ref{sec_2}, we formulate a MFSM to determine the $t^*(\boldsymbol{x})$ field based on the $t(\boldsymbol{x})$ field. In this section, we validate the MFSM by comparing it with analytical solutions and analyze its errors as the grid size is adjusted, considering several models with different velocity and attenuation arrangements. According to eq. \eqref{tstar_eik}, it is evident that the velocity and attenuation, along with the traveltime field $t$ determined by velocity (eq. \eqref{t_eik}), affect $t^*$. In the following numerical tests, we separately arrange the models with homogeneous and heterogeneous attenuation structures while varying velocity models to investigate the characteristics of the $t^*(\boldsymbol{x})$ field,  accompanied by comparing it with the traveltime field $t(\boldsymbol{x})$ in the Cartesian and spherical coordinates. Noted that these models are constructed in 3D; unless otherwise specified, the third dimension consists of only three grid nodes.


\subsection{Uniform velocity and attenuation model}\label{sec_3_1}
We begin by considering Model 1, as shown in Figs \ref{fig:model1}(a) and (b), which is characterized by uniform $V_{P}$ and $Q_{P}$ in Cartesian coordinates. In this model, $V_{P}$ is set as 4.0 km/s, and $Q_{P}$ is specified as 500.0. The model extends 30 km in both the $x$ and $z$ directions. The source, indicated by the red star, is positioned at (15 km, 15 km). The region is discretized using spatial intervals of $\Delta x$ = $\Delta z$ = 0.20 km, resulting in a grid of 151 by 151 points. 

The black solid line in Fig. \ref{fig:model1}(a) represents the isolines of $t(\boldsymbol{x})$ calculated using the FSM, while the blue solid line in Fig. \ref{fig:model1}(b) represents the isolines of $t^*(\boldsymbol{x})$ calculated using the MFSM. For better illustration, the $t(\boldsymbol{x})$ and $t^*(\boldsymbol{x})$ fields, computed numerically, along with their corresponding isolines, are also presented in Figs. \ref{fig:model1}(c) and (d), respectively. For the uniform model, the analytical solutions for $t(\boldsymbol{x})$ and $t^*(\boldsymbol{x})$ can be easily derived. The isolines of the $t$ and $t^*$ fields calculated using the analytical solution are represented by the black dashed lines in Figs. \ref{fig:model1}(a) and (c), and the blue dashed lines in Fig. \ref{fig:model1}(b) and (d), respectively. We observe strong consistency between the $t(\boldsymbol{x})$ values calculated by the FSM and their analytical counterparts, as well as between the $t^*(\boldsymbol{x})$ values calculated by the MFSM and their analytical solutions (Figs \ref{fig:model1}a-d). The pattern of the $t^*(\boldsymbol{x})$ isolines closely resembles that of the $t(\boldsymbol{x})$ isolines. However, the values represented by the isolines differ, with $t^*(\boldsymbol{x})$ being proportional to $t$ by a factor of $1/500.0$. This proportionality is evident from eqs \eqref{t_eik} and \eqref{tstar_eik}, where the only difference is that $t^*$ can be obtained by multiplying both sides of eq. \eqref{t_eik} by $1/Q_{P}$, assuming a uniform $Q_{P}$.

Figs \ref{fig:model1}(e) and (f) show the absolute errors in the $t(\boldsymbol{x})$ and $t^*(\boldsymbol{x})$ fields, comparing the results obtained using the FSM or MFSM with those derived from the analytical solutions. We can observe that, for both $t$ and $t^*$, the absolute errors are smallest (equal to zero) along the principal axes (horizontal and vertical directions relative to the source) and largest along the diagonal directions. This behavior is attributed to the angle between wavefront propagation and grid orientation, which is zero along the principal axes and reaches $45^{\circ}$ in the diagonal directions. The larger the angle, the greater the errors introduced when solving eqs \eqref{t_eik} and \eqref{tstar_eik} using the FSM and MFSM with a rectangular mesh. Beyond the principal axes, the absolute errors in $t$ and $t^*$ accumulate with increasing distance from the source (Figs \ref{fig:model1}e and f). A similar phenomenon is observed in the grid-based fast marching method (FMM) when solving the traveltime field $t$ \citep{alkhalifah2001}. The relative errors (in percentage) for $t(\boldsymbol{x})$ and $t^*(\boldsymbol{x})$ are presented in Figs \ref{fig:model1}(g) and (h). For both $t$ and $t^*$, the percentage errors are larger near the source decrease significantly with distance. This occurs because, in Cartesian coordinates with a regular grid distribution, the wavefront curvature near the source is high and undersampled, whereas further from the source, the wavefront becomes flatter and is oversampled \citep[e.g.,][]{alkhalifah2001,lan2012,lan2013topo,zhou2023topo}. Close to the source (Figs \ref{fig:model1}g-h), the relative errors in $t$ and $t^*$ accumulates dramatically due to the singularity at the point source \citep[e.g.,][]{alkhalifah2001,rawlinson2004,fomel2009fast}.

Although the uniform model is the simplest, it provides valuable insights into the fundamental characteristics of the $t^*(\boldsymbol{x})$ field. The accuracy of $t^*$ is influenced by factors such as the angle between wavefront propagation and grid orientation, the curvature of the wavefront relative to the grid size, and the singularity at the point source. 


\begin{figure}
\center{\includegraphics[width=16cm]{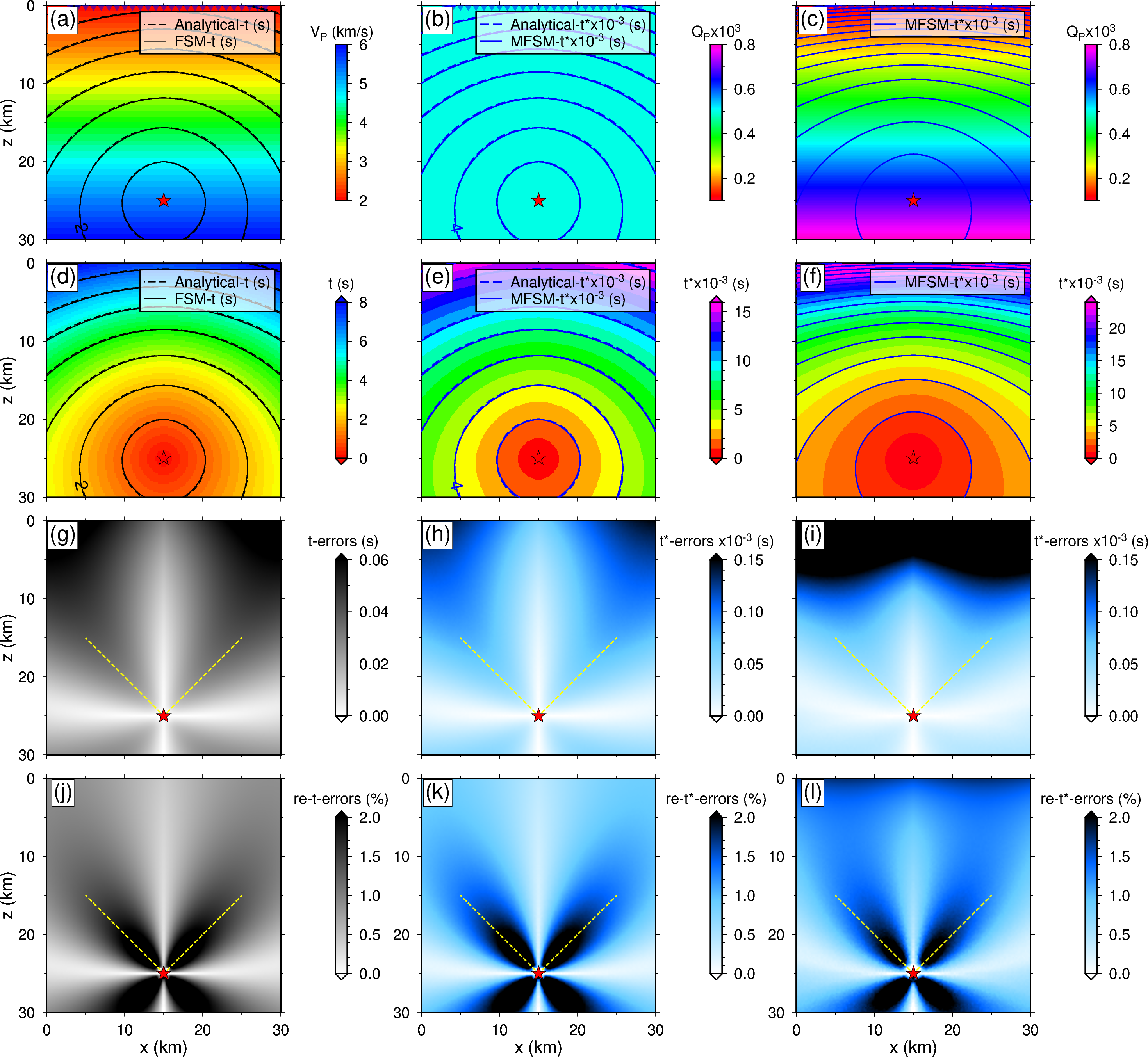}}
\caption
{Model 2 consists of $V_{P}$ model with a constant-gradient (a) and $Q_{P}$ model either uniform (b) or with a constant-gradient (c). The velocity model in (a) is represented by $V_{P}(\boldsymbol{x})$ = $V_{P}(x, y, z)$ = 2.0 + $\frac{30z}{6.0-2.0}$. The $Q_{P}$ value in (b) is 500.0, while in (c), it is represented by $Q_{P}(\boldsymbol{x})$ = $Q_{P}(x, y, z)$ = 100.0 + $\frac{30z}{800.0-100.0}$. The red star represents the point source, while the blue inverted triangles in (a), (b), and (c) indicate a horizontal receiver array. The region is discretized with a 0.20 km $\times$ 0.20 km grid, resulting in 151 $\times$ 151 points. The comparison of isolines for $t(\boldsymbol{x})$  from the numerical and analytical solutions is shown in (a) and (d), with the absolute and relative (in percentage) errors presented in (g) and (j), respectively. With the inclusion of the uniform $Q_{P}$ model (b), the comparison of isolines for $t^*(\boldsymbol{x})$ from the numerical and analytical solutions is shown in (b) and (e), with the absolute and relative (in percentage) errors presented in (h) and (k), respectively. With the inclusion of the constant-gradient $Q_{P}$ model (c), the isolines for $t^*(\boldsymbol{x})$ is shown in (c) and (f), while the absolute and relative (in percentage) errors between grid sizes of 0.20 km and 0.01 km are presented in (i) and (l), respectively. The yellow dashed lines represent auxiliary lines along the diagonal directions.}
\label{fig:model2_1}
\end{figure}



\begin{figure}
\center{\includegraphics[width=15cm]{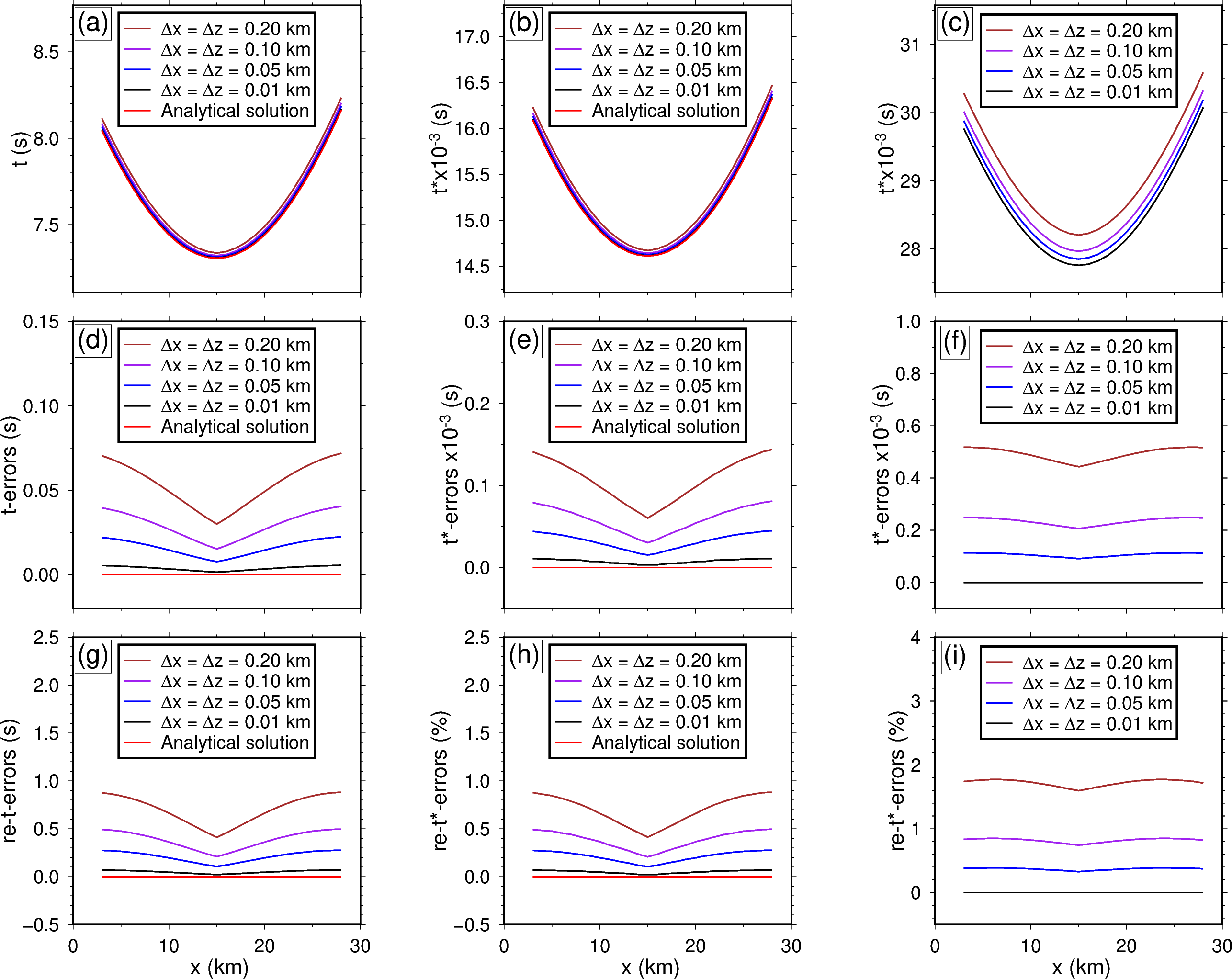}}
\caption
{Values of $t$ and $t^*$ recorded by a horizontal receiver array in Model 2, calculated for both the analytical solution (red solid line) and the numerical solution with grid refinement. The receiver array spans from (3 km, 0 km) to (28 km, 0 km). Compared to the analytical solution, the $t$ values, along with the absolute and relative $t$ errors (in percentage) calculated using the FSM with grid refinement are shown in (a), (d) and (g), respectively. With the inclusion of the uniform $Q_{P}$ model (Fig. \ref{fig:model2_1}b), (b), (e) and (h) display the $t^*$ values, along with the absolute and relative $t^*$ errors (in percentage) calculated by the MFSM with grid refinement, compared to the analytical solution. With the inclusion of the constant-gradient $Q_{P}$ model (Fig. \ref{fig:model2_1}c), (c), (f) and (i) display the $t^*$ values, along with the absolute and relative $t^*$ errors (in percentage) calculated using the MFSM with grid refinement, compared to the solutions obtained with the dense grid of a size of 0.01 km.}
\label{fig:model2_2}
\end{figure}

\subsection{Constant-gradient velocity and attenuation models}\label{sec_3_2}

To further validate the MFSM for solving $t^*$, we now consider models that incorporate constant-gradient velocity and attenuation. Model 2 consists of a velocity model with a constant-gradient, $V_{P}(\boldsymbol{x})$ = $V_{P}(x, y, z)$ = 2.0 + $\frac{30z}{6.0-2.0}$ (Fig. \ref{fig:model2_1}a). This is combined with an attenuation model that either has a uniform $Q_{P} = 500.0$ (Fig. \ref{fig:model2_1}b) or a constant-gradient $Q_{P}(\boldsymbol{x})$ = $Q_{P}(x, y, z)$ = 100.0 + $\frac{30z}{800.0-100.0}$ (Fig. \ref{fig:model2_1}c). The source, marked by the red star, is located at (15 km, 25 km). As in Section \ref{sec_3_1}, the region is partitioned with spatial intervals of $\Delta x$ = $\Delta z$ = 0.20 km, producing a grid of 151 by 151 points.



In Fig. \ref{fig:model2_1}(a), as $V_{P}$ decreases from 6.0 km/s to 2.0 km/s in the upward direction, $t(\boldsymbol{x})$ isolines become denser gradually. With the uniform $Q_{P}$ model applied, the pattern of $t^*$ isolines (Fig. \ref{fig:model2_1}b) matches that of the $t$ isolines (Fig. \ref{fig:model2_1}a), with their values maintaining a consistent ratio of 1/500.0. However, when incorporating the constant-gradient $Q_{P}$ model, the gradual decrease in $Q_{P}$ in the upward direction leads to much denser $t^*$ isolines (Fig. \ref{fig:model2_1}c). The analytical solution for $t(\boldsymbol{x})$ in a constant-gradient velocity model can be found in \cite{fomel2009fast}. For a uniform $Q_{P}$ model (Figs. \ref{fig:model2_1}a and b), the derivation of the analytical solution for $t^*(\boldsymbol{x})$ is straightforward. However, when both $V_{P}$ and $Q_{P}$ have constant gradients (Figs. \ref{fig:model2_1}a and c), obtaining the analytical solution for $t^*(\boldsymbol{x})$ becomes challenging. In Figs \ref{fig:model2_1}(a)-(b), the comparisons between the analytical solutions (dashed lines) and the numerical results from the FSM for $t(\boldsymbol{x})$ (black solid lines), as well as from the MFSM for $t^*(\boldsymbol{x})$ (blue solid lines), exhibit good consistency. In Fig. \ref{fig:model2_1}(c), only the $t^*$ values calculated by the MFSM are shown, as no analytical solution exist for the constant-gradient attenuation model. The $t(\boldsymbol{x})$ and $t^*(\boldsymbol{x})$ fields, along with their corresponding isolines, are shown in Figs. \ref{fig:model2_1}(d)-(f).

The absolute $t(\boldsymbol{x})$ and $t^*(\boldsymbol{x})$ errors are illustrated in Figs \ref{fig:model2_1}(g)-(i), while the relative errors are shown in Figs \ref{fig:model2_1}(j)-(l). Compared to Figs \ref{fig:model1}(e) and (f), Figs \ref{fig:model2_1}(g) and (h) show that the presence of a gradient in $V_{P}$ leads to different behaviors in the absolute $t$ and $t^*$ errors. For instance, the absolute $t$ and $t^*$ errors directly above the source are no longer zero. Similar phenomena are observed in the relative errors shown in Figs \ref{fig:model2_1}(j) and (k). The main reason for these phenomena is the inexact first-order approximation of the derivatives of $t$ with respect to $z$, as $V_{P}$ changes gradually in the $z$ direction. Another observation is that the largest absolute and relative errors for $t$ and $t^*$ (Figs. \ref{fig:model2_1}g-h and j-k) progressively deviate from the diagonal directions as depth decreases. We attribute this deviation to the decrease in $V_{P}$ with decreasing depth. As discussed in Section \ref{sec_3_1}, the primary source of error stems from propagation distance inaccuracies caused by the angle between wavefront propagation and grid orientation. Thus, in the constant-gradient velocity model (Fig. \ref{fig:model2_1}a), shallower depths with lower $V_{P}$ result in larger $t$ and $t^*$ errors. For the uniform $Q_{P}$ model (Fig. \ref{fig:model2_1}b), variations in the absolute and relative $t^*$ errors result from changes in $V_{P}$ and inaccuracies in $t$, with the latter also being influenced by variations in $V_{P}$. For the constant-gradient $Q_{P}$ model (Fig. \ref{fig:model2_1}c), the results calculated using a dense grid with a size of 0.01 km are taken as the reference solution. As $Q_{P}$ decreases from 800.0 to 100.0 in the upward direction, the absolute and relative $t^*$ errors (Fig \ref{fig:model2_1}i and l) also deviate from the diagonal directions at shallower depths. This can be attributed to the fact that lower $Q_{P}$ values at shallower depths are more likely to generate larger $t^*$ errors.

To analyze the errors more quantitatively, Fig. \ref{fig:model2_2} presents the records of a horizontal array of receivers positioned at locations from (3 km, 0 km) to (28 km, 0 km). Additionally, we present the results calculated with grid sizes refined from 0.2 km to 0.1 km, 0.05 km, and 0.01 km. For the constant-gradient $V_{P}$ and uniform $Q_{P}$ model, $t$ and $t^*$ calculated with different grid sizes show strong consistency with their analytical solutions (Figs \ref{fig:model2_2}a-b). However, for the constant-gradient $V_{P}$ and constant-gradient $Q_{P}$ model, noticeable differences in $t^*$ can be observed when using different grid sizes (Fig \ref{fig:model2_2}c). The absolute and relative errors in $t$ and $t^*$ decrease systematically as the grid is refined (Figs \ref{fig:model2_2}d-i). Specifically, for the uniform $Q_{P}$ model (Fig. \ref{fig:model2_1}b), refining the grid size from 0.2 km to 0.01 km reduces the maximum absolute $t^*$ error from $\sim$ $0.144\times10^{-3}$ s (brown solid line) to $\sim$ $0.011\times10^{-3}$ s (black solid line) (Fig. \ref{fig:model2_2}e). Similarly, the maximum relative $t^*$ error decreases from $\sim$ 8.8 \text{\textperthousand} (brown solid line) to $\sim$ 0.7 \text{\textperthousand} (black solid line) in Fig. \ref{fig:model2_2}(h). For the constant-gradient $Q_{P}$ model (Fig. \ref{fig:model2_1}c), as the grid size decreases from 0.2 km to 0.01 km, the maximum absolute $t^*$ error decreases by $\sim$ $0.518\times10^{-3}$ s (brown and black solid lines) (Fig. \ref{fig:model2_2}f), while the maximum relative $t^*$ error decreases by $\sim$ 1.77 $\%$ (brown and black solid lines) (Fig. \ref{fig:model2_2}i). At a grid size of 0.01 km (black solid line), the calculated results show negligible deviations from the analytical solutions (Figs. \ref{fig:model2_2}e and h). Thus, we use the solutions calculated with this grid size as the reference for the constant-gradient $V_{P}$ and constant-gradient $Q_{P}$ model (Figs. \ref{fig:model2_2}f and i).

\begin{figure}
\center{\includegraphics[width=15cm]{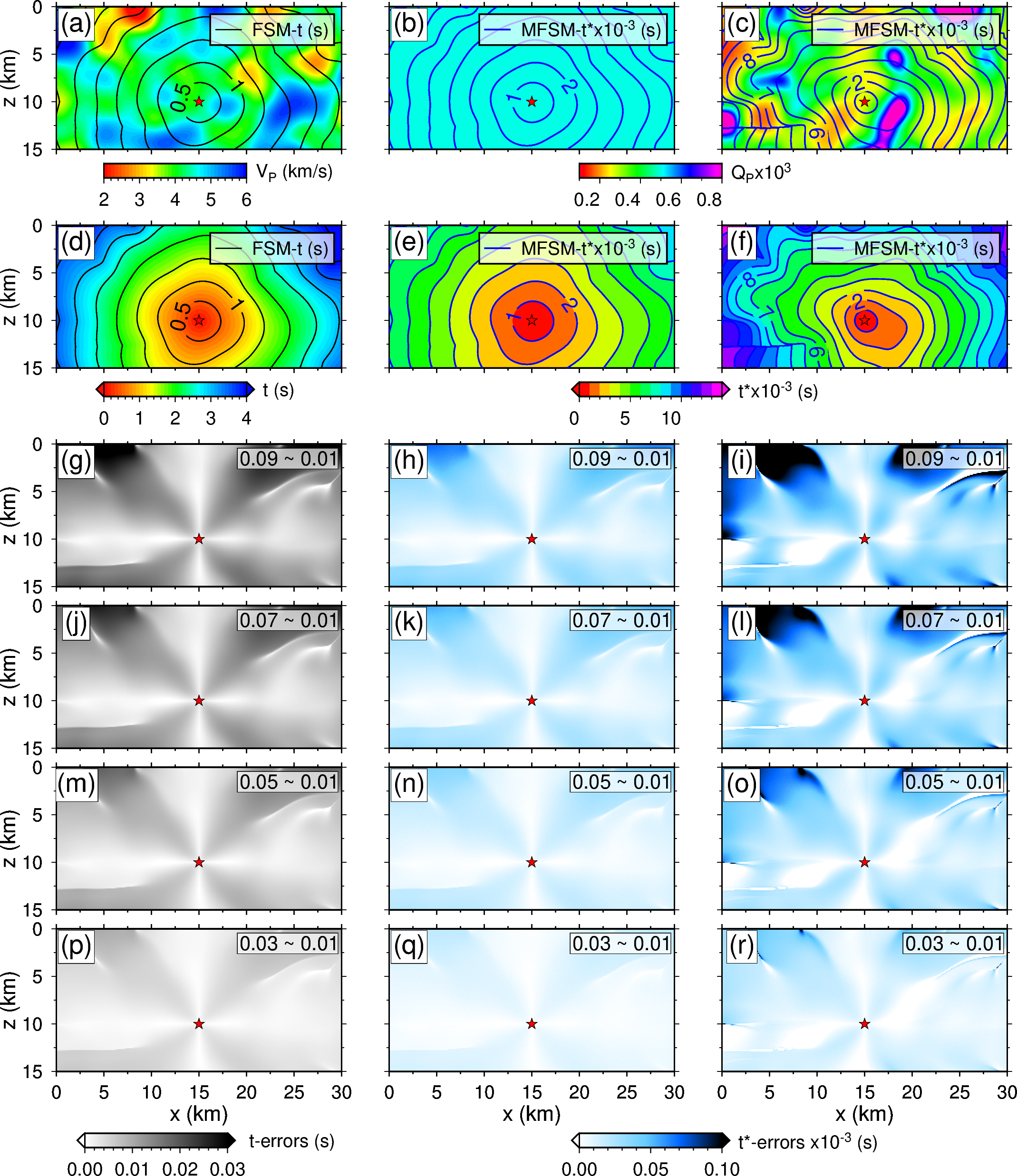}}
\caption
{Model 3 features a $V_{P}$ model with heterogeneous anomalies (a) and a $Q_{P}$ model either uniform (b) or with heterogeneous anomalies (c). The source (red star) is positioned at (15 km, 10 km). The isolines for $t(\boldsymbol{x})$ (black solid lines) are shown in (a) and (d). With the inclusion of the uniform $Q_{P}$ model (b), the isolines of $t^*(\boldsymbol{x})$ (blue solid lines) are shown in (b) and (e). With the inclusion of the heterogeneous $Q_{P}$ model (c), the isolines of $t^*(\boldsymbol{x})$ (blue solid lines) are presented in (c) and (f). The $t(\boldsymbol{x})$ and $t^*(\boldsymbol{x})$ fields are shown in (d), (e) and (f), respectively. With grid refinement from 0.09 km to 0.01 km, subgraphs (g), (j), (m) and (p) display the absolute $t(\boldsymbol{x})$ errors, and subgraphs (h), (k), (n) and (q) show the absolute $t^*(\boldsymbol{x})$ errors for the uniform $Q_{P}$ model, and subgraphs (i), (l), (o) and (r) show the absolute $t^*(\boldsymbol{x})$ errors for the heterogeneous $Q_{P}$ model. All errors are referenced against the dense grid of a size of 0.01 km.}
\label{fig:model4}
\end{figure}

Variations in $V_{P}$ and/or $Q_{P}$ influence the accumulation of $t^*$ errors, with lower values of $V_{P}$ and $Q_{P}$ generally leading to larger errors. However, as the grid size is refined, these $t^*$ errors progressively decrease and approach the analytical solution. These results demonstrate the effectiveness of the MFSM in accurately solving $t^*$ in models with constant-gradient velocity and attenuation. 


\subsection{Heterogeneous velocity and attenuation models}\label{sec_3_3}

In this section, we validate the MFSM for computing $t^*$ in complex heterogeneous models with varying $V_{P}$ and $Q_{P}$, both in Cartesian and spherical coordinates.

\subsubsection{Examples in Cartesian coordinates}\label{sec_3_3_1}

We examine Model 3 in Cartesian coordinates, where the heterogeneous $V_{P}$ model is randomly generated (Fig \ref{fig:model4}a), superimposed with an attenuation model that is either a uniform ($Q_{P} = 500.0$, Fig \ref{fig:model4}b) or randomly generated (Fig \ref{fig:model4}c). The modeled region spans 30 km in the $x$ direction and 15 km in the $z$ direction. The source is placed at (15 km, 10 km). The region is discretized on grids with sizes $\Delta x = \Delta z$ of 0.09 km, 0.07 km, 0.05 km, 0.03 km and 0.01 km, corresponding to grid resolutions of 335$\times$168, 430$\times$216, 601$\times$301, 1001$\times$501 and 3001$\times$1501, respectively. 

The isolines of $t(\boldsymbol{x})$ and $t^*(\boldsymbol{x})$ are shown as black and blue solid lines in Figs \ref{fig:model4}(a), (b) and (c). Additionally, the $t(\boldsymbol{x})$ and $t^*(\boldsymbol{x})$ fields, along with their isolines, are presented in Figs. \ref{fig:model4}(d), (e), and (f). Since the high and low $V_{P}$ and $Q_{P}$ anomalies are randomly distributed across the region, the isolines of $t(\boldsymbol{x})$ and $t^*(\boldsymbol{x})$ exhibit heterogeneous patterns. With a uniform $Q_{P}$ model (Fig \ref{fig:model4}b), the $t^*(\boldsymbol{x})$ field and its isolines closely resemble the patterns of the $t(\boldsymbol{x})$ field and its isolines. However, when the random $Q_{P}$ model is applied (Fig \ref{fig:model4}c), the $t^*(\boldsymbol{x})$ field and its corresponding isolines become much more heterogeneous (Fig \ref{fig:model4}f). The solutions computed using the finest grid size of 0.01 km are taken as the reference solution. The absolute errors in $t(\boldsymbol{x})$ and $t^*(\boldsymbol{x})$ for different grid sizes are illustrated in Figs \ref{fig:model4}(g)–(r). The patterns of these errors are closely linked to the distribution of the heterogeneous $V_{P}$ and $Q_{P}$ anomalies. Although the magntitudes of the $t(\boldsymbol{x})$ and $t^*(\boldsymbol{x})$ errors vary across the region, they systematically decrease as the grid size is refined from 0.09 km to 0.01 km (Figs \ref{fig:model4}g–r).

\subsubsection{Numerical Examples in spherical coordinates}\label{sec_3_3_2}
\begin{figure}
\center{\includegraphics[width=14.0cm]{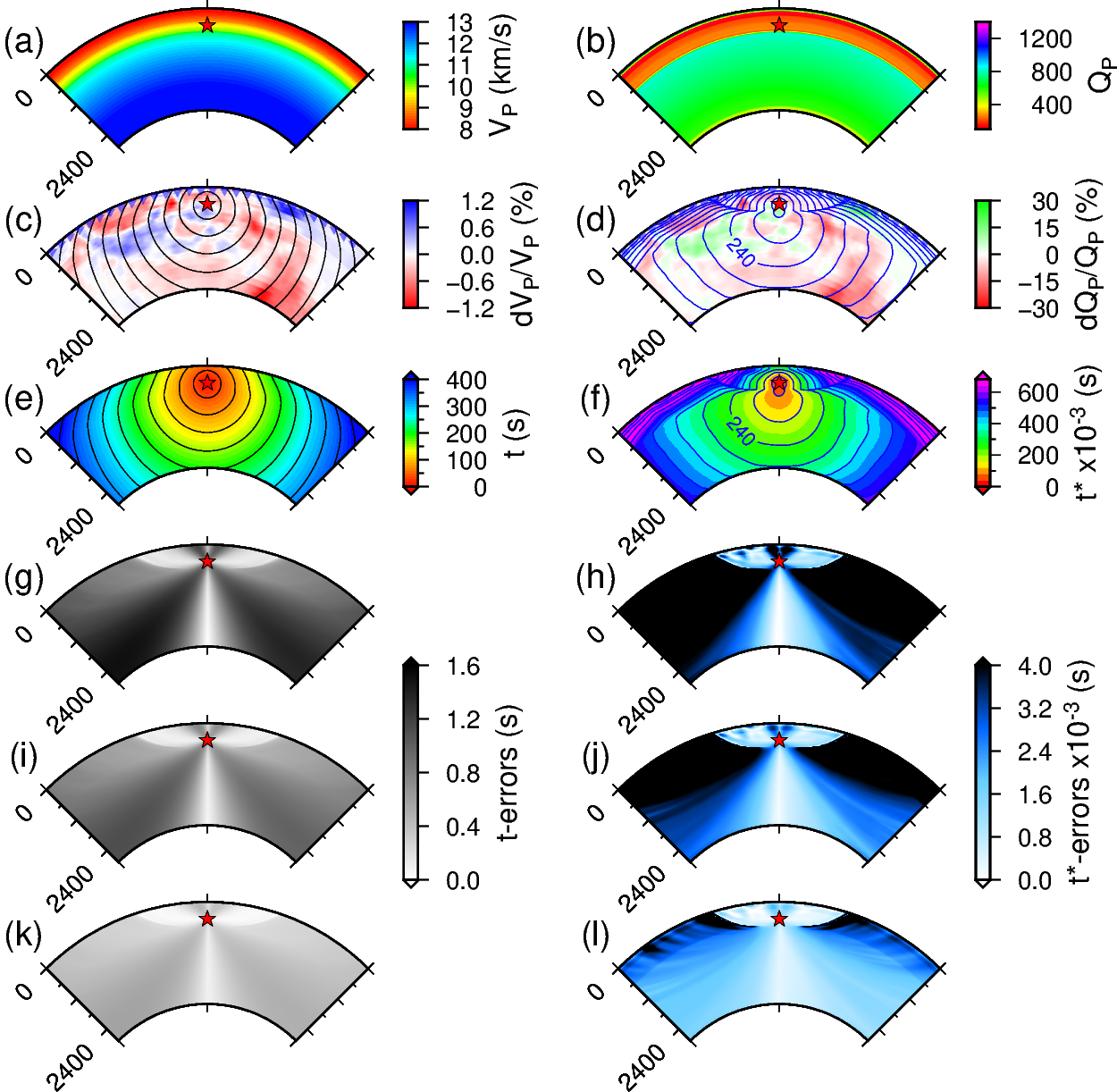}}
\caption
{Model 4 consists of a $V_{P}$ model showing the percent deviation (c) from the ak135 reference $V_{P}$ model (a) and a $Q_{P}$ model showing the percent deviation (d) from the attenuation model (b) derived from the ak135 reference. The source (red star) is positioned at 500.0 km depth. The isolines of $t(\boldsymbol{x})$ (black solid lines) are shown in (c) and (e), and the isolines of $t^*(\boldsymbol{x})$ (blue solid lines) are displayed in (d) and (f). The $t(\boldsymbol{x})$ and $t^*(\boldsymbol{x})$ fields are illustrated in (e) and (f), respectively. The blue inverted triangles in (c)-(f) indicate the locations of a receiver array. With grid refinement from 0.004 rad and 4.0 km, 0.003 rad and 3.0 km, 0.002 rad and 2.0 km to 0.001 rad and 1.0 km, the absolute $t(\boldsymbol{x})$ errors are presented in (e), (g), (i) and (k), while the absolute $t^*(\boldsymbol{x})$ errors are shown in (f), (h), (j) and (l), all referenced against the dense grid of 0.001 rad and 1.0 km.}
\label{fig:model5_1}
\end{figure}

To broaden the method's applicability, we extend its application to compute $t^*$ in spherical coordinates. The theoretical framework for this extension is detailed in Appendix A. For Model 4, the $V_{P}$ and $Q_{P}$ models are based on the AK135 reference model \citep{kennet1995mod}, as shown in Figs \ref{fig:model5_1}(a) and (b). It should be noted that $Q_{P}$ is determined by $Q_{\mu}$ and $Q_{\kappa}$, as no $Q_{P}$ model is available in the Ak135 model. The $V_{P}$ perturbation, depicted in Fig. \ref{fig:model5_1}(c), is derived from the MITP2008 model \citep{li2008}, and represents the percentage deviation of $V_{P}$ from the AK135 reference model. The $Q_{P}$ perturbation, shown in Fig. \ref{fig:model5_1}(d), is obtained by scaling the $V_{P}$ perturbation from Fig. \ref{fig:model5_1}(c). The source, marked by a black star, is positioned at a depth of 500.0 km.

\begin{figure}
\center{\includegraphics[width=15.0cm]{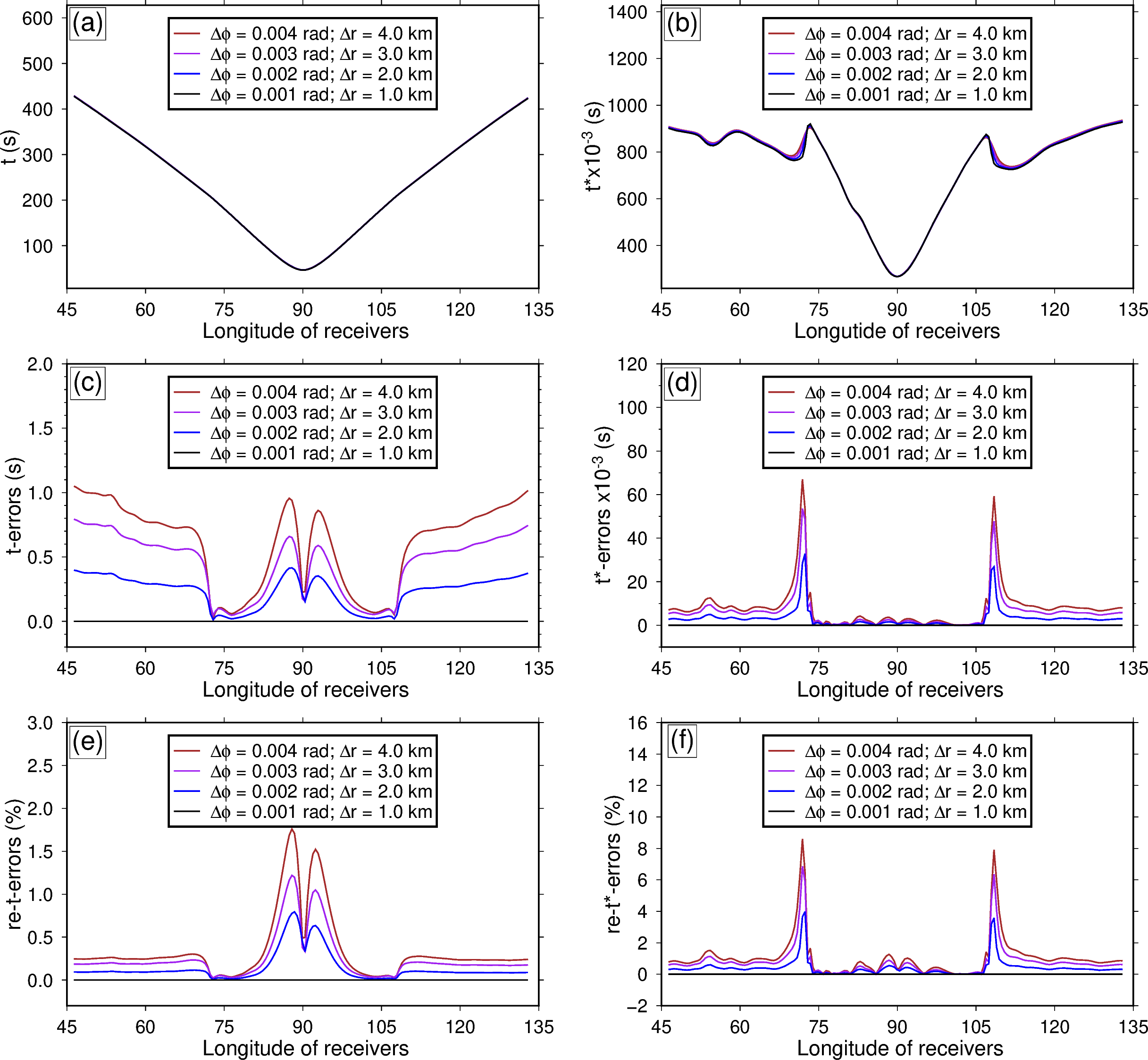}}
\caption
{(a) and (b) present the values of $t$ and $t^*$, respectively, recorded by a horizontal receiver array in Model 4, as calculated by the FSM and the MFSM with grid refinement. (c) and (e)  show the absolute and relative $t$ errors (in percentage), respectively, calculated by the FSM with grid refinement, while (d) and (f) display the absolute and relative $t^*$ errors (in percentage), respectively, calculated by the MFSM with grid refinement. All errors are referenced to the dense grid of 0.001 rad and 1.0 km.}
\label{fig:model5_2}
\end{figure}

With the grid size adjusted to $\Delta \phi$ = 0.004 rad, 0.003 rad, 0.002 rad, and 0.001 rad, and $\Delta r$ = 4.0 km, 3.0 km, 2.0 km, and 1.0 km, the grid resolutions are $390 \times 712$, $520 \times 949$, $780 \times 1424$, and $1559\times2847$, respectively. It should be noted that, while the calculations are conducted in 3D spherical coordinates ($r$, $\theta$, $\phi$), only three grid nodes are arranged in the $\theta$-direction. The isolines of $t$ and $t^*$, calculated using the FSM and MFSM, are shown in Figs \ref{fig:model5_1}(c) and (d), respectively. Additionally, the $t(\boldsymbol{x})$ and $t^*(\boldsymbol{x})$ fields, along with their isolines, are presented in Figs \ref{fig:model5_1}(e) and (f). Absolute $t(\boldsymbol{x})$ and $t^*(\boldsymbol{x})$ errors for varying grid sizes are illustrated in Figs \ref{fig:model5_1}(g)–(l). These errors, particularly the $t^*(\boldsymbol{x})$ errors, decrease significantly as the grid is refined. We also present $t$ and $t^*$ records for a receiver array positioned at the surface, spanning from ($45^\circ$, 0 km) to ($135^\circ$, 0 km) in Fig. \ref{fig:model5_2}. For different grid sizes, the $t$ and $t^*$ values remain consistent (Figs \ref{fig:model5_2}a and b). The solutions for $t$ and $t^*$ calculated with the finest grid ($\Delta \phi$ = 0.001 rad and $\Delta r$ = 1.0 km) are used as reference solutions. As shown in Figs \ref{fig:model5_2}c-f, both the absolute and relative errors of $t$ and $t^*$ decrease as the grid is refined. Specifically, as the grid size changes from $\Delta \phi$ = 0.004 rad and $\Delta r$ = 4.0 km to $\Delta \phi$ = 0.001 rad and $\Delta r$ = 1.0 km, the maximum absolute $t^*$ errors decrease by $67.0 \times 10^{-3}$ s (brown and black solid lines) (Fig \ref{fig:model5_2}d), and the relative $t^*$ errors decrease by $8.59 \%$ (brown and black solid lines) (Fig \ref{fig:model5_2}f).


These results, combined with those presented in Section \ref{sec_3_3_1}, demonstrate that the MFSM is effective in solving for $t^*$ in models with heterogeneous $V_{P}$ and $Q_{P}$, in both Cartesian and spherical coordinates.

\section{Convergence Analysis, Iteration Count and Computational time}\label{sec_4}
We use Model 3 and Model 4 to quantitatively analyze the convergence behavior, iteration count, and computational time with grid refinement. The mean $L^{1}$ error for $t^*$ over the entire region is calculated using the following formula:
\begin{equation}\label{l1_error}\begin{split}
&L^{1}(t^*) = \frac{{\sum^{M_{I}}_{i=1}}{\sum^{M_{J}}_{j=1}}{\sum^{M_{K}}_{k=1}} \left | t^*_{ref,\,i,\,j,\,k} - t^*_{num,\,i,\,j,\,k}\right |}{M_{I}M_{J}M_{K}}.
\end{split}\end{equation}
where both $t^*_{ref}$ and $t_{num}$ are numerical solutions computed by the MFSM. Note that $t^*_{ref}$ is the numerical solution computed with the finest mesh and $t_{num}$ is the numerical solution computed with other mesh sizes. $M_{I}$, $M_{J}$ and $M_{K}$, represent the number of grid points along the three coordinate axes, as introduced in Section \ref{sec_2_2}. The order of accuracy for the $n$-th grid is calculated by
\begin{equation}\label{l1_error}\begin{split}
&\mathrm{order} = \frac{\mathrm{ln}\frac{[L^{1}(t^*)]_{n}}{[L^{1}(t^*)]_{n-1}}}{\mathrm{ln}\frac{h_{n}}{h_{n-1}}},
\end{split}\end{equation}
where $[L^{1}(t^*)]_{n}$ and $h_{n}$ are the $L^{1}(t^*)$ error and grid spacing for the $n$-th grid, respectively. 

For Model 3, the grid spacing, number of gridpoints, $L^{1}(t^*)$ error, order of accuracy, and computing time for various mesh sizes are shown in Table \ref{model4}. The reference solution is computed using a grid spacing of 0.01 km. As the grid spacing is refined from 0.10 km to 0.02 km, the $L^{1}$ errors decreased significantly, from $3.999 \times 10^{-5}$ s to $0.599 \times 10^{-5}$ s. We can observe that with grid refinement, the MFSM achieves first-order accuracy and, in some cases, even higher accuracy (Table \ref{model4}). The convergence criterion for calculating $t^*$, defined as the maximum difference between two consecutive iterations, is set to $\upsilon = 1.0 \times 10^{-12}$ s. For Model 3,  only three Gauss-Seidel iterations are required to meet the convergence criterion. Similarly, for Model 4, Table \ref{model5} summarizes the grid spacing, number of gridpoints, $L^{1}(t^*)$ error, order of accuracy, and computing time for different mesh sizes. The reference solution is computed using a grid spacing of $\Delta{\phi} = 0.001$ rad and $\Delta{r} = 1.0$ km. The $L^{1}$ errors decrease from $44.127 \times 10^{-4}$ s to $17.636 \times 10^{-4}$ s as the grid is refined from $\Delta{\phi} = 0.004$ rad and $\Delta{r} = 4.0$ km to $\Delta{\phi} = 0.002$ rad and $\Delta{r} = 2.0$ km. The MFSM method can achieve precision exceeding first order when solving for $t^*$ in spherical coordinates. Notably, only two Gauss-Seidel iterations are required to meet the convergence criterion. This analysis further manifests that the developed MFSM is accurate, efficient, and exhibits robust convergence in calculating $t^*$ in both Cartesian and spherical coordinates system.


\begin{table*}
\begin{minipage}{150mm}
\caption{Error analysis and computational efficiency of the MFSM for Model 3 in Cartesian coordinates.}
\label{model4}
\begin{tabular}{@{}cccccc}
Grid size ($\Delta{x};\Delta{y};\Delta{z}$) & Mesh & $L^{1}$ errors ($\times 10^{-5}$ s) & Order of accuracy & Time (s) & Iterations\\
\hline
0.10 km  & 301$\times$151$\times$3  & 3.999  &  -  & 0.516  & 3 \\
0.09 km  & 335$\times$168$\times$3  & 3.536  &  1.170  & 0.526  & 3 \\
0.08 km  & 376$\times$189$\times$3  & 3.212  &  0.816  & 0.779  & 3 \\
0.07 km  & 430$\times$216$\times$3  & 2.873  &  0.835  & 0.980  & 3 \\
0.06 km  & 501$\times$251$\times$3  & 2.470  &  0.979  & 1.305  & 3 \\
0.05 km  & 601$\times$301$\times$3  & 2.051  &  1.019  & 2.010  & 3 \\
0.04 km  & 751$\times$376$\times$3  & 1.583  &  1.161  & 3.069  & 3 \\
0.03 km  & 1001$\times$501$\times$3  & 1.125  & 1.187  & 5.750  & 3 \\
0.02 km  & 1501$\times$751$\times$3  & 0.599  &  1.556  & 15.062  & 3 \\
0.01 km  & 3001$\times$1501$\times$3  & -  &  -  & 74.111  & 3 \\
\hline
\end{tabular}
\end{minipage}
\end{table*}

\begin{table*}
\begin{minipage}{150mm}
\caption{Error analysis and computational efficiency of the MFSM for Model 4 in spherical coordinates.}
\label{model5}
\begin{tabular}{@{}cccccc}
Grid size ($\Delta{\phi};\Delta{\theta};\Delta{r}$) & Mesh & $L^{1}$ errors ($\times 10^{-4}$ s) & Order of accuracy & Time (s) & Iterations\\
\hline
0.004 rad; 4.0 km  & 390$\times$712$\times$3  & 44.127  & -  & 2.719  & 2 \\
0.003 rad; 3.0 km  & 520$\times$949$\times$3  & 33.106  &  0.999  & 5.201  & 2 \\
0.002 rad; 2.0 km  & 780$\times$1424$\times$3  & 17.636  &  1.553  & 12.078  & 2 \\
0.001 rad; 1.0 km  & 1559$\times$2847$\times$3  & -  &  -  & 64.011  & 2 \\
\hline
\end{tabular}
\end{minipage}
\end{table*}

\section{Application in North Island, New Zealand}\label{sec_5}
\begin{figure}
\center{\includegraphics[width=16.0cm]{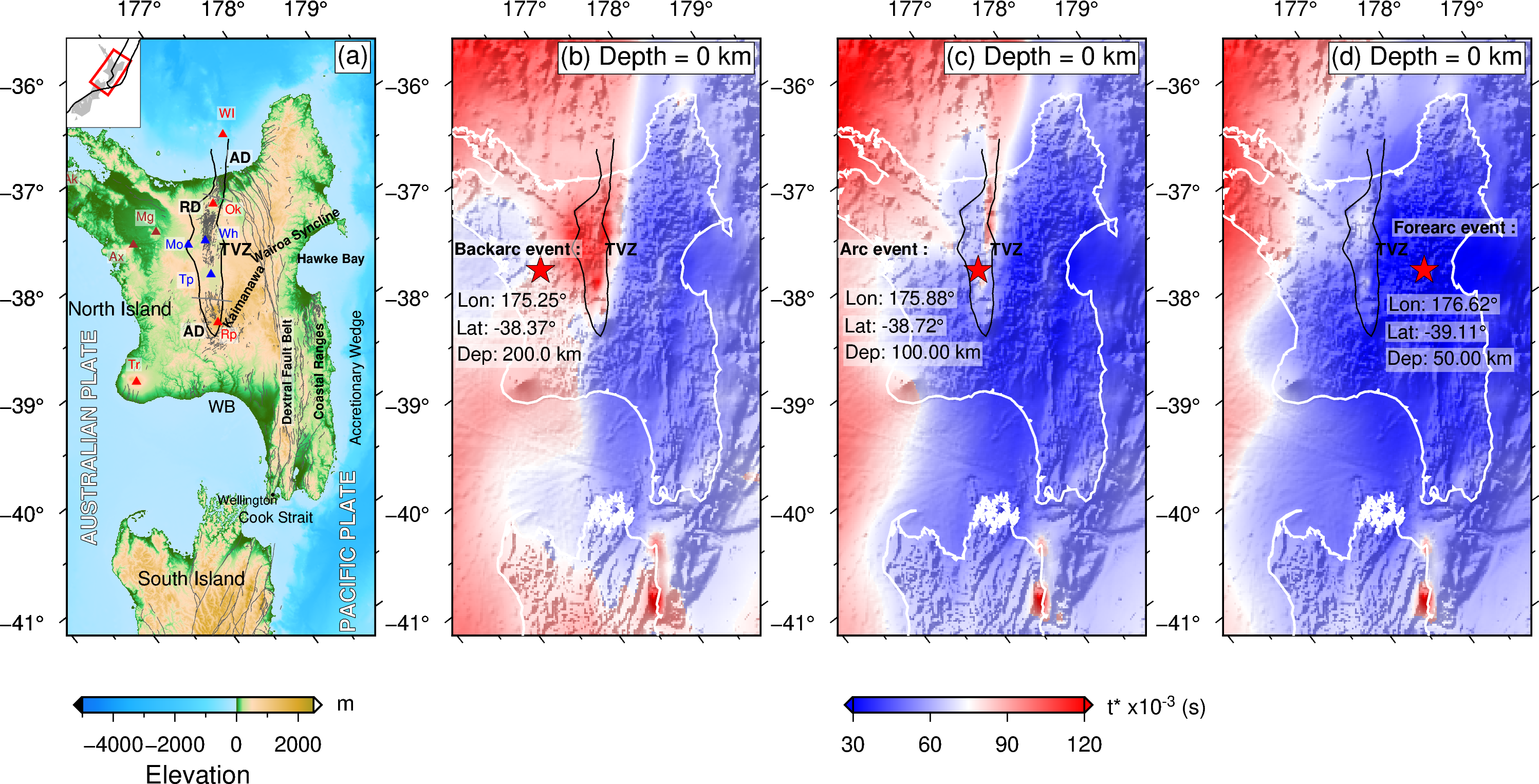}}
\caption
{(a) Map of North Island, New Zealand, showing topography and major tectonic features. Major faults are indicated by gray lines. The research region is outlined by the red solid lines in the inset at the upper left corner, where the boundaries of the Australian Plate and the Pacific Plate are marked by black solid lines. The Taupo Volcanic Zone (TVZ) is marked by black solid curves. Volcanic features are marked as follows: calderas (blue triangles), active andesite volcanoes (red triangles), and non-silicic volcanic centers (brown triangles). Volcanic labels: Ak, Auckland; Ax, Alexandra; Mg, Maungataurari; WI, White Island; Ok, Okataina; Rp, Ruapehu; Tr, Taranaki; Mo, Mangakino; Wh, Whakamaru; Tp, Taupo. (b), (c) and (d) show surface $t^*$ maps at 0 km depth, with slab earthquakes (red star) located beneath the backarc, arc, and forearc regions, respectively.}
\label{fig:nnz_1}
\end{figure}

In this section, we use the MFSM to solve for the $t^*$ field based on the realistic $V_{P}$ and $Q_{P}$ models of the central North Island (Fig. \ref{fig:nnz_1}a), and discuss the application of $t^*$ in estimating earthquake response spectra.

The topographic map of the central North Island is shown in (Fig. \ref{fig:nnz_1}a). The main tectonic and geological setting is as follows: In the central North Island, the Pacific and Australian plates are converging obliquely at a rate of 42 mm $yr^{-1}$ \citep{demets1994}. This convergence is accommodated by the subduction of the Pacific plate and deformation of the overlying Australian plate. The North Island is situated above the Hikurangi subduction zone, which has resulted in volcanism and extension in the Taupo Volcanic Zone (TVZ). Volcanic activity within the TVZ varies along strike, with rhyolite-dominated caldera volcanoes in the central section and andesite-dominated cone volcanoes to the north and south (Fig. \ref{fig:nnz_1}a).

\begin{figure}
\center{\includegraphics[width=14cm]{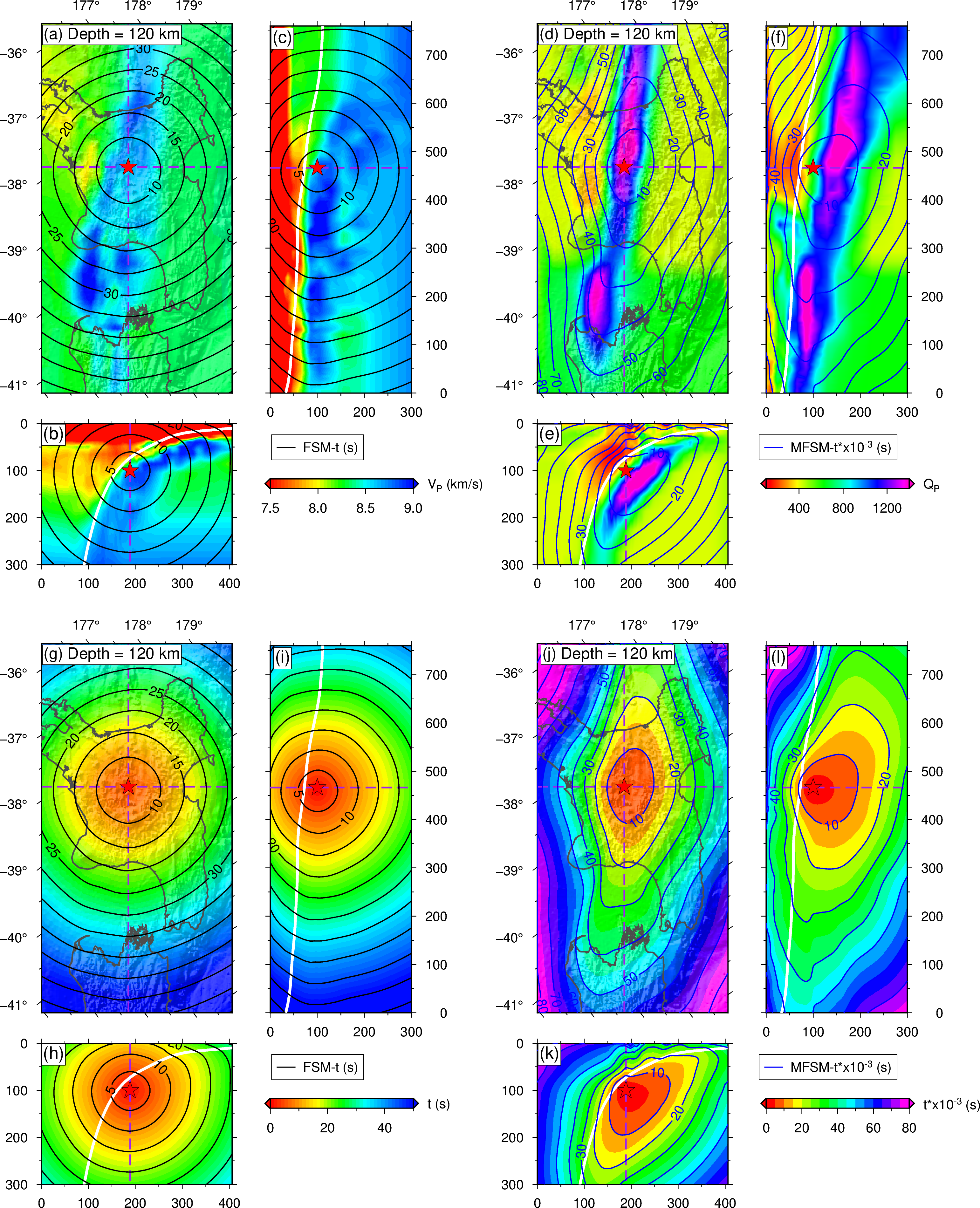}}
\caption
{The horizontal and vertical sections passing through the source (red star) of the $V_{P}$ (a-c) and $Q_{P}$ (d-f) models, along with the $t(\boldsymbol{x})$ isolines (a-c), calculated using the FSM and $t^*(\boldsymbol{x})$ isolines (d-f), calculated by the MFSM. The $t(\boldsymbol{x})$ and $t^*(\boldsymbol{x})$ fields, along with their corresponding isolines, are shown in panels (g)-(i) and (j)-(l), respectively. The earthquake, located beneath the arc region (Fig \ref{fig:nnz_1}b), has coordinates of $175.88^{\circ}$ longitude, $-38.72^{\circ}$ latitude, and a depth of $100.0$ km. The white solid lines in each vertical cross-section represent the location of the Slab 2.0 interface.}
\label{fig:nnz_2}
\end{figure}

The research region, outlined by the red solid line in the inset of Fig. \ref{fig:nnz_1}(a), has an oblique shape. To reduce computational costs, the region is transformed into a regular grid with coordinates spanning $[-1.7^{\circ},\enspace2.0^{\circ}] \times [-3.0^{\circ},\enspace4.0^{\circ}]$. The 3-D $V_{P}$ and $Q_{P}$ models are extracted from the New Zealand Wide Model 2.2 \citep[e.g.,][]{eberhart2010,eberhart2015} and interpolated onto a grid with dimensions of $160\times300\times160$. In Fig. \ref{fig:nnz_2}, horizontal and vertical sections of the $V_{P}$ and $Q_{P}$ models passing through the source (red star) located at coordinates ($175.88^{\circ}$, $-38.72^{\circ}$, $100.0$ km) are presented. The primary characteristics are as follows. The velocity model shows high $V_{P}$ for the slab relative to the mantle wedge and reduced velocity for the mantle below the TVZ (Figs \ref{fig:nnz_2}a-c). The $Q_{P}$ model exhibits high $Q_{P}$ ($900-1200$) for the subducted cold slab and low $Q_{P}$ ($< 400$) for the mantle wedge (Figs \ref{fig:nnz_2}d-f). Significant variations are observed in the mantle wedge along the strike of the subduction zone, with the most pronounced low $Q_{P}$ ($< 250$) occurring in the mantle wedge between depths of 50 and 85 km beneath the rhyolite-dominated, productive central segment of the Taupo Volcanic Zone (Figs \ref{fig:nnz_2}e-f). 


We consider three slab earthquakes at depths ranging from 50.0 km to 200.0 km and calculate their corresponding $t^*$ fields. As the slab earthquake depth is adjusted from 200.0 km to 100.0 km and then to 50.0 km, it transitions from beneath the backarc, through the arc, to the forearc. As previously mentioned, $t^*$ characterizes the spectral attenuation of earthquakes (magnitude $\geq$ 2.5) \citep{eberhart2003} through $e^{-{{\omega}{t^*}/2}}$ \citep[e.g.,][]{cormier1982,bindi2006}, which is routinely recorded by the seismographic network at the surface. Accordingly, the $t^*$ maps at a depth of 0 km for these three earthquakes are shown in Figs \ref{fig:nnz_1}(b)-(d). For earthquakes at different depths, the corresponding $t^*$ maps show distinct patterns (Figs \ref{fig:nnz_1}b-d). For the earthquake beneath the backarc (Fig. \ref{fig:nnz_1}b), a pronounced high $t^*$ is observed at the TVZ, while a reduction in $t^*$ appears northwest of the TVZ. For earthquakes beneath the arc and forearc, the high $t^*$ region progressively shifts from southeast to northwest, with these regions showing progressively lower $t^*$ (Figs \ref{fig:nnz_1}c-d). Notably, in Figs \ref{fig:nnz_1}(c) and (d), the TVZ consistently shows relatively high $t^*$ compared to the surrounding regions. To explain these features, we present horizontal and vertical sections of the $t$ and $t^*$ fields for the earthquake beneath the arc in Fig. \ref{fig:nnz_2}. Compared to the $t$ isolines shown in Figs \ref{fig:nnz_2}(a)-(c) and (g)-(i), the $t^*$ isolines are highly heterogeneous (Figs \ref{fig:nnz_2}d-f and j-l). The low $Q_{P}$ in the crust and mantle wedge beneath the TVZ compresses the $t^*$ isolines, resulting in larger $t^*$ (Fig. \ref{fig:nnz_1}c). Conversely, the low $t^*$ in the fore-arc region can mainly be attributed to wave propagation through the high $Q_{P}$ slab.

The significant variations in $V_{P}$ and $Q_{P}$ across the North Island result in distinct $t^*$ maps at the surface for earthquakes occurring at different depths (Figs \ref{fig:nnz_1}b-d). These synthetic $t^*$ values for various earthquakes can be converted into a path-averaged attenuation rate ($CQ$), which can then be used to refine the standard response spectral attenuation model \citep[e.g.,][]{eberhart2003,eberhart2010b}. To estimate $CQ$, it is usually necessary to calculate $t^*$ for a range of source-receiver pairs. This is conventionally done by integrating along ray paths using $t^* =\int_{L}{1}/({V(\boldsymbol{x})Q(\boldsymbol{x})})dl$. However, as mentioned earlier, calculating $t^*$ using this integration-based method can be time-consuming, especially when numerous source-receiver pairs are involved \citep{rawlinson2004}. As shown in Figs \ref{fig:nnz_1}(b)-(d), the developed MFSM, which eliminates the need for ray tracing, provides a convenient to obtain $t^*$, thereby facilitating the estimation of earthquake response spectra.

\section{Discussion}\label{sec_6}
The traditional FSM \citep{zhao2005fast} is used to solve the eikonal and  $t^*$-governing equations, but it does not account for the point source singularity, leading to inaccuracies near the source. Factored techniques \citep[e.g.,][]{fomel2009fast,luo2012,luo2014} have been developed to address this issue by better approximating spherical wavefronts near the source. Incorporating these techniques will improve the accuracy of $t$ and $t^*$ calculations.

Moreover, the FSM and MFSM are tested on a planar Earth's surface, but real-world applications often involve complex topography, such as mountainous or volcanic regions \citep[e.g.,][]{sun2011,garcia2012,prudencio2015anta,prudencio2015cana}. By integrating methods like unstructured grids \citep[e.g.,][]{rawlinson2004,qian2007fast} or topography-dependent formulations \citep[e.g.,][]{lan2013topo,zhou2023topo}, the MFSM could be adapted to effectively calculate $t$ and $t^*$ on irregular surfaces.

The fast marching method (FMM), another widely used finite-difference approach, solves the eikonal equation by tracking monotonically advancing wavefronts \citep[e.g.,][]{zhao2005fast,qian2007fast,luo2012}. Given the dependence of $t^*$ on the gradient of $t$, $t^*$ could be obtained by solving its governing equation while using the FMM to solve for $t$. This could potentially be developed into a modified FMM to directly solve $t^*$ without using ray tracing. Incorporating alternative methods like the fast iterative method (FIM) \citep[e.g.,][]{jeong2008}, finite-element method (FEM) \citep[e.g.,][]{pullan2002}, or discontinuous Galerkin method (DGM) \citep[e.g.,][]{cheng2007} into $t^*$ solutions could expand forward modeling tools, particularly for adjoint-state attenuation tomography \citep[e.g.,][]{huang2020,he2020}.

\section{Conclusions}
Seismic attenuation (mainly intrinsic and scattering attenuation) can cause changes in amplitude and phase for a propagating seismic wave, with these changes characterized by the attenuation operator $t^*$. Traditionally, $t^*$ is calculated by integrating the inverse of the velocity-quality factor $Q$ product along a ray path determined via ray tracing. To avoid ray tracing and enable the development of a ray-free attenuation tomography method, we propose a modified fast sweeping method (MFSM) to directly compute $t^*$.

The MFSM computes $t^*$ by numerically solving the differential form of its governing equation. First, the integral expressions for $t$ and $t^*$ are converted into differential forms using directional derivatives. By relating the directional derivative to the gradient, the eikonal equation for $t$ and the governing equation for $t^*$ are derived. The traveltime $t$ is numerically computed using the classic fast sweeping method \citep{zhao2005fast}, and it remains fixed while solving for $t^*$. Given that $t^*$ and $t$ share the same ray path, the gradient of $t^*$ (${\nabla t^*}$) is discretized using the upwinding scheme derived from ${\nabla t}$. The $t^*$ field is then calculated by solving the discretized $t^*$ governing equation using the Fast Sweeping Algorithm, based on the velocity and attenuation models, and the determined $t$ field.

The developed MFSM is validated through several numerical experiments. First, $t^*$ values calculated by the MFSM are compared with analytical solutions for uniform and constant-gradient models, with an analysis of the absolute and relative $t^*$ errors. For heterogeneous velocity and attenuation models, $t^*$ values are calculated with grid refinement, and the convergence, iteration count, and computation time of the MFSM are also evaluated. The effectiveness of the MFSM is demonstrated in both Cartesian and spherical coordinates. Finally, using a realistic velocity and attenuation model for North Island, New Zealand, we compute the $t^*$ field for different slab earthquakes and discuss the surface $t^*$ values and their role in estimating earthquake response spectra. In future work, this MFSM will be applied to develop an adjoint-state attenuation tomography method. 



\begin{acknowledgments}
This study is funded by Minister of Education, Singapore, under its MOE AcRF Tier-2 Grant (MOE-T2EP20122-0008)
\end{acknowledgments}

\begin{dataavailability}
\end{dataavailability}
All data produced in this work are available upon request via email at \href{dongdong.wang@ntu.edu.sg}{dongdong.wang@ntu.edu.sg}. The velocity and attenuation model for the North New Zealand are publicly available from NZ-Wide2.2 (\href{https://zenodo.org/records/3779523}{https://zenodo.org/records/3779523}).


\appendix \label{app_A}
\section{The Calculation of traveltime $t$ and attenuation operator $t^*$ in Spherical coordinates}
To address the specific application scenario for the earthquake tomography, we present the algorithm for solving eqs \eqref{t_eik} and \eqref{tstar_eik} in spherical coordinates $\boldsymbol{x}=(r,\theta,\phi)$. $r$ represents the distance to the Earth's centre. $\theta$ and $\phi$ respectively denote the latitude and longitude. The gradient of $t(\boldsymbol{x})$ is $\nabla t(\boldsymbol{x})=(\partial_{r}t,\frac{1}{r}\partial_{\theta}t,\frac{1}{r\mathrm{cos}{\theta}}\partial_{\phi}t)$. $\Omega$ represents the Earth's volume in the 3D space $\mathcal{R} \times \Theta \times \Phi = [0, \infty) \times [-\frac{1}{2}\pi,\frac{1}{2}\pi] \times[0,2\pi)$. In spherical coordinates, the computational domain $\Omega$ is partitioned into a uniform mesh with grid points $\boldsymbol{x}_{i,j,k}$ and mesh sizes  $\Delta r$, $\Delta \theta$ and $\Delta \phi$. The total number of the grid points in three directions are $M_{I}$, $M_{J}$ and $M_{K}$, respectively. Employing the Godunov upwind difference scheme to discretize eq. \eqref{t_eik} at interior grid points $(2 \leq i \leq M_{I}-1,2 \leq j \leq M_{J}-1,2 \leq k \leq M_{K}-1)$:
\begin{equation}\label{t_disc_sp}
[\frac{(t_{i,j,k}-t^{r\, \mathrm{min}}_{i,j,k})^+}{\Delta r}]^2+
[\frac{(t_{i,j,k}-t^{\theta\, \mathrm{min}}_{i,j,k})^+}{r_{i,j,k} \Delta \theta}]^2 + 
[\frac{(t_{i,j,k}-t^{\phi\, \mathrm{min}}_{i,j,k})^+}{r_{i,j,k}{\mathrm{cos}}{{\theta}_{i,j,k}} \Delta \phi}]^2 
= s_{i,j,k}^2
\end{equation}
where
\begin{equation}\label{t_min_sp}
t^{r\, \mathrm{min}}_{i,j,k} = \mathrm{min}(t_{i-1,j,k},t_{i+1,j,k}),t^{\theta\, \mathrm{min}}_{i,j,k} = \mathrm{min}(t_{i,j-1,k},t_{i,j+1,k}),t^{\phi\, \mathrm{min}}_{i,j,k} = \mathrm{min}(t_{i,j,k-1},t_{i,j,k+1})
\end{equation}
and
\begin{equation}\label{A3}\begin{split}
(x)^+ = \left\{
      \begin{aligned}
      x, \: x \enspace \textgreater 0,\\
      0, \: x \leq 0.
      \end{aligned}
\right.
\end{split}\end{equation}
At the boundary of the domain (i.e., $i=1\lor M_{I}$;$j=1\lor M_{J}$;$k=1\lor M_{K}$), the one-sided differences are used. For example, at the upper boundary $\boldsymbol{x}_{1,j,k}$, we have
\begin{equation}\label{t_disc_sp_one}
[\frac{(t_{1,j,k}-t_{2,j,k})^+}{\Delta r}]^2+
[\frac{(t_{1,j,k}-t^{\theta\, \mathrm{min}}_{i,j,k})^+}{r_{i,j,k} \Delta \theta}]^2 + 
[\frac{(t_{1,j,k}-t^{\phi\, \mathrm{min}}_{i,j,k})^+}{r_{i,j,k}{\mathrm{cos}}{{\theta}_{i,j,k}} \Delta \phi}]^2 
= s_{i,j,k}^2.
\end{equation}
Similar to that used in the Cartesian coordinates, we then employ the Fast Sweeping Algorithm to solve eqs \eqref{t_disc_sp} and \eqref{t_disc_sp_one}.

In spherical coordinates, eq. \eqref{tstar_eik} governing $t^{*}$ can be written in the following difference form,
\begin{equation}\label{tstar_disc_sp}\begin{split}
&\frac{(t_{i,j,k}-t^{r\, \mathrm{min}}_{i,j,k})^+}{\Delta r}\frac{(t^*_{i,j,k}-t^{*,\, r\, \mathrm{min}}_{i,j,k})}{\Delta r}+
\frac{(t_{i,j,k}-t^{\theta\, \mathrm{min}}_{i,j,k})^+}{r_{i,j,k} \Delta \theta}\frac{(t^*_{i,j,k}-t^{*,\, \theta\, \mathrm{min}}_{i,j,k})}{r_{i,j,k} \Delta \theta} +\\& 
\frac{(t_{i,j,k}-t^{\phi\, \mathrm{min}}_{i,j,k})^+}{r_{i,j,k}{\mathrm{cos}}{{\theta}_{i,j,k}} \Delta \phi}\frac{(t^*_{i,j,k}-t^{*,\, \phi\, \mathrm{min}}_{i,j,k})}{r_{i,j,k}{\mathrm{cos}}{{\theta}_{i,j,k}} \Delta \phi} 
= s_{i,j,k}^2 q_{i,j,k},
\end{split}\end{equation}
where $t^{*,\, r\, \mathrm{min}}_{i,j,k}$, $t^{*,\, \theta\, \mathrm{min}}_{i,j,k}$ and $t^{*,\, \phi\, \mathrm{min}}_{i,j,k}$ can be determined by:
\begin{equation}\label{A3-1}\begin{split}
t^{*,\, r\, \mathrm{min}}_{i,j,k} = \left\{
      \begin{aligned}
      t^*_{i-1,j,k}, \: \enspace \textup{if} \enspace t^{r\, \mathrm{min}}_{i,j,k} = t_{i-1,j,k} ,\\
      t^*_{i+1,j,k}, \: \enspace \textup{if} \enspace {t^{r\, \mathrm{min}}_{i,j,k} = t_{i+1,j,k}} ,
      \end{aligned}
\right.
\end{split}\end{equation}

\begin{equation}\label{A3-2}\begin{split}
t^{*,\, \theta\, \mathrm{min}}_{i,j,k} = \left\{
      \begin{aligned}
      t^*_{i,j-1,k}, \: \enspace \textup{if} \enspace t^{{\theta}\, \mathrm{min}}_{i,j,k}= t_{i,j-\rm{1},k}  ,\\
      t^*_{i,j+1,k}, \: \enspace \textup{if} \enspace t^{{\theta}\, \mathrm{min}}_{i,j,k} = t_{i,j+1,k},
      \end{aligned}
\right.
\end{split}\end{equation}

\begin{equation}\label{A3-3}\begin{split}
t^{*,\, \phi\, \mathrm{min}}_{i,j,k} = \left\{
      \begin{aligned}
      t^*_{i,j,k-1}, \: \enspace \textup{if} \enspace t^{{\phi}\, \mathrm{min}}_{i,j,k} = t_{i,j,k-1},\\
      t^*_{i,j,k+1}, \: \enspace \textup{if} \enspace t^{{\phi}\, \mathrm{min}}_{i,j,k} = t_{i,j,k+1}.
      \end{aligned}
\right.
\end{split}\end{equation}
While employing one-sided differences at the boundary of the computational region, the difference form at the upper boundary, for example, will be
\begin{equation}\label{tstar_disc_sp_one}\begin{split}
&\frac{(t_{1,j,k}-t_{2,j,k})^+}{\Delta r}\frac{(t^*_{1,j,k}-t^*_{2,j,k})}{\Delta r}+
\frac{(t_{1,j,k}-t^{\theta\, \mathrm{min}}_{i,j,k})^+}{r_{i,j,k} \Delta \theta}\frac{(t^*_{1,j,k}-t^{*,\, \theta\, \mathrm{min}}_{i,j,k})}{r_{i,j,k} \Delta \theta} +\\& 
\frac{(t_{1,j,k}-t^{\phi\, \mathrm{min}}_{i,j,k})^+}{r_{i,j,k}{\mathrm{cos}}{{\theta}_{i,j,k}} \Delta \phi}\frac{(t^*_{1,j,k}-t^{*,\, \phi\, \mathrm{min}}_{i,j,k})}{r_{i,j,k}{\mathrm{cos}}{{\theta}_{i,j,k}} \Delta \phi} 
= s_{i,j,k}^2 q_{i,j,k},
\end{split}\end{equation}
Finally, we use the Fast Sweeping Algorithm to solve eqs \eqref{tstar_disc_sp} and \eqref{tstar_disc_sp_one}.

\end{document}